\documentclass[12pt,a4paper]{iopart}

\usepackage{colordvi}
\usepackage{graphicx,sidecap}
\usepackage{bm}
\usepackage{braket}
\usepackage{amssymb}
\usepackage{mathrsfs}
\usepackage{amsmath}
\usepackage{xcolor}
\usepackage{hyperref}
\usepackage{enumerate}
\usepackage[toc,title]{appendix}
\usepackage{verbatim}
\usepackage[margin=30pt, bf, font=footnotesize, center, justification=justified]{caption}[2004/07/16]
\hypersetup{colorlinks,bookmarksopen,bookmarksnumbered,
 citecolor=red,
 linkcolor=blue,
 pdfstartview=green,
 urlcolor=purple}

\newcommand{\eq}[1]{\begin{equation}#1\end{equation}}

\newcommand{\ee}{\mathrm{e}}

\def\be{\begin{equation}}
\def\ee{\end{equation}}
\def\bea{\begin{eqnarray}}
\def\eea{\end{eqnarray}}

\begin{document}

\title[Entanglement Hamiltonians for non-critical quantum chains]
{Entanglement Hamiltonians for non-critical quantum chains}

\author{ Viktor Eisler$^1$, Giuseppe Di Giulio$^2$, Erik Tonni$^2$ and \\ Ingo Peschel$^3$}

\address{$^1$Institut f\"ur Theoretische Physik, Technische Universit\"at Graz, Petersgasse 16, A-8010 Graz, Austria}
\address{$^2$SISSA and INFN Sezione di Trieste, via Bonomea 265, I-34136 Trieste, Italy}
\address{$^3$Fachbereich Physik, Freie Universit\"at Berlin, Arnimallee 14, D-14195 Berlin, Germany}

\begin{abstract}
We study the entanglement Hamiltonian for finite intervals in infinite
quantum chains for two different free-particle systems: coupled harmonic
oscillators and fermionic hopping models with dimerization. Working in the
ground state, the entanglement Hamiltonian describes again free bosons or
fermions and is obtained from the correlation functions via high-precision
numerics for up to several hundred sites. Far away from criticality, the
dominant on-site and nearest-neighbour terms have triangular profiles 
that can be understood from the analytical results for a half-infinite
interval. Near criticality, the longer-range couplings, although small,
lead to a more complex picture. 
A comparison between the exact spectra and entanglement
entropies and those resulting from the dominant terms in the Hamiltonian
is also reported. 
\end{abstract}
\maketitle
%%%%%%%%%%%%%%%%%%%%%%%%%%%%%%%%%%%%%%%%%%%%%%%%%%%%%%%%%%%%%%%%%%%%%%%%%%%%%%%%%%%%%%%%%%%%%
\section{Introduction}
\label{sec:intro}

To study the entanglement properties of quantum systems, one divides the full system into
two parts and determines how they are coupled in the chosen state
\cite{Calabrese/Cardy/Doyon09,Peschel/Eisler09,Casini/Huerta09,Eisert/Cramer/Plenio10}.
This information is encoded in the reduced density matrix $\rho$ of one of the pieces, and this quantity
can always be written in the form $\rho=\exp(-\mathcal{H})/Z$. One therefore is dealing
with a kind of statistical mechanics problem, but the operator $\mathcal{H}$, called the
entanglement Hamiltonian, depends on the quantum state in question as well as on the type
of partition and differs in general from the physical Hamiltonian of the subsystem.

For chains in their ground state, the simplest case is an infinite system divided into
two half-infinite ones. Then $\mathcal{H}$ is an operator in which the terms increase linearly
as one moves away from the interface. For continuous critical systems, this result is attributed
to Bisognano and Wichmann \cite{Bisognano/Wichmann75,Bisognano/Wichmann76} and described
by the formula
\begin{equation}
 \mathcal{H}=2\pi \int_0^{\infty}dx\,x\,\mathcal{T}_{00}(x)
\label{BW}
\end{equation}
where $\mathcal{T}_{00}(x)$ is the energy density in the physical Hamiltonian. This formula already
contains the essence of the situation: the operator $\mathcal{H}$ describes an inhomogeneous
system with small terms near the boundary and large ones in the interior of the subsystem.
This also holds for the non-critical case, both in the continuum and on a lattice. In the
latter case it follows, for integrable chains, from the relation of $\rho$ to corner transfer
matrices (CTMs) \cite{Nishino/Okunishi97,Peschel/Kaulke/Legeza99,Peschel/Eisler09} and the
particular structure of these matrices first noted by Baxter \cite{Baxter76,Baxter77,Baxter82}.
Roughly speaking,
the linear increase reflects the geometrical widening of the annular sections in the associated
two-dimensional partition functions as the distance from the corner increases.

For other partitions and geometries of continuous critical systems, the form of $\mathcal{H}$
can be obtained from conformal invariance \cite{Hislop/Longo82,Casini/Huerta/Myers11,Wong_etal13,Cardy/Tonni16}.
For example, a subsystem in the form of an interval of length $\ell$ in an infinite chain leads to
\begin{equation}
 \mathcal{H}=2\pi \ell \int_0^{\ell}dx\,\beta(x)\,\mathcal{T}_{00}(x)
\label{interval}
\end{equation}
where the parabola $\beta(x)=x/\ell(1-x/\ell)$ increases linearly at \emph{both} ends of the
interval. This has been checked in various numerical calculations. For free fermions
on a lattice, one finds that $\mathcal{H}$ contains nearest-neighbour hopping which does not
quite vary parabolically and, in addition, hopping to more distant neighbours with smaller
amplitudes \cite{Peschel/Eisler09,Eisler/Peschel17}. However, it has been shown numerically
\cite{Arias_etal17_1} and also analytically \cite{Eisler/Tonni/Peschel19} that by properly
including the longer-range terms in the continuum limit one recovers the conformal
result for $\beta(x)$. The same was found for free bosons in the form of coupled harmonic
oscillators \cite{DiGiulio/Tonni19}. Some results also exist for small intervals in interacting
fermion systems \cite{Nienhuis/Campostrini/Calabrese09,Parisen/Assaad18}.

The goal of the present work is to characterize $\mathcal{H}$ for an interval in chains
\emph{away} from criticality, and we do this by studying two free-particle models which are generalizations
of those just mentioned. For the bosons, the frequency $\omega$ of each single oscillator is kept
finite, while for the fermions a dimerization is introduced via alternating hopping matrix elements
$t(1 \pm \delta)$. This corresponds to the Su-Schrieffer-Heeger model for polyacetylene in the
absence of interactions \cite{Su/Schrieffer/Heeger79,Heeger/Kivelson/Schrieffer88}. In both cases,
the ground states have Gaussian nature and $\mathcal{H}$ is a free-particle Hamiltonian which can be
determined from the correlation functions of the chains \cite{Peschel03,Peschel/Eisler09}. This is
done with high-precision numerics which allows to treat large intervals. Both chains can also 
be related to integrable two-dimensional models which leads to explicit formulae for $\mathcal{H}$
if the interval is half-infinite.

We find that the basic pattern is always similar to the critical case: there are some dominant
terms in $\mathcal{H}$ whereas all others are much smaller. In the bosonic case, these are
the diagonal matrix elements in the kinetic and in the potential energy and the nearest-neighbour
coupling in the latter, while in the fermionic model at half filling, it is the nearest-neighbour
hopping. These quantities vanish linearly at the ends of the interval, and the linear behaviour
extends more and more into the interior, as one moves away from criticality. In the end, the
curves approach a triangular form instead of a parabola. This corresponds to a combination of
the effects from the two boundaries, and the slopes are given correctly by the CTM results for
the half-infinite subsystem. Defining an approximate entanglement Hamiltonian with these dominant
terms, one finds that, except at the upper end, its spectrum is identical to that of the true one.
Therefore, it also gives the same entanglement entropy except very close to criticality.
These features are completely analogous to those found for critical chains with a parabolic
variation of the couplings in $\mathcal{H}$ \cite{Giudici/etal18,Giudici/etal19,ZCDR20}.

In the fermionic case, there is an additional feature due to the dimerization: the dimerization
pattern of the physical Hamiltonian is found again in the entanglement Hamiltonian, as already
noted in \cite{Eisler/Chung/Peschel15}. The even and odd bonds differ, and this is particularly
marked in the centre, but both show a trend towards a triangular variation as the dimerization
increases. In contrast to the critical case, however, an operator constructed from them commutes
only approximately with the entanglement Hamiltonian.

The behaviour of the small matrix elements, which describe longer-range couplings, is more complex.
They show spatial oscillations which are absent at criticality and can vanish outside a region around
the middle of the interval. A particular subset corresponds to couplings across the centre.
The region where they have relatively large values has its maximal extent when the correlation length
is comparable to the size of the interval. Because of these features we were not able to obtain
a consistent continuum picture near the critical point.

The layout of the paper is as follows. In section \ref{sec:setting}, we describe the setting and give
the basic formulae, in particular for the elements of the correlation functions.
Section \ref{sec:halfchain} presents explicit expressions for the entanglement Hamiltonians of
half-infinite subsystems, which serve as points of reference for the case of an interval.
In section \ref{sec:osc}, the numerical results for the elements in $\mathcal{H}$ are presented for
intervals in strongly non-critical oscillator chains and a simplified version of $\mathcal{H}$ is discussed.
In section \ref{sec:dim}, the same is done for the dimerized hopping chain.
Section \ref{sec:genfeat} is devoted to the general features of $\mathcal{H}$ in the non-critical region,
including long-range couplings across the middle of the interval, while section \ref{sec:summary}
sums up our findings and also addresses the question of a continuum limit.
Finally, in appendices \ref{app:ent} and \ref{app:half-line-continuum}, the entanglement entropy and the
continuum form of $\mathcal{H}$ are derived for a half-infinite subsystem of the oscillator chain,
while appendix \ref{app:comm} discusses a fermionic operator which almost commutes with $\mathcal{H}$.

%%%%%%%%%%%%%%%%%%%%%%%%%%%%%%%%%%%%%%%%%%%%%%%%%%%%

\section{Setting\label{sec:setting}}

In this section, we describe the two chains we shall study and give
the formulae from which the entanglement Hamiltonian $\mathcal{H}$ follows.
For a subsystem of $N$ sites, its diagonal form reads
\eq{
\mathcal{H} = \sum_{l=1}^{N} \varepsilon_l \, f_l^{\dag} f^{\phantom{\dag}}_l 
\label{ehdiag}}
where $f_l^{\dag}$ and $f_l$ are either bosonic or fermionic
creation and annihilation operators and $\varepsilon_l$ denote
the single-particle eigenvalues.
They are determined via elementary correlation matrices
restricted to the given segment in the quantum chain at hand,
with the relation depending on the particle statistics.
In order to obtain the entanglement Hamiltonian in real space,
the operators $f_l^{\dag}$ and $f_l$ have to be transformed back into the original
variables, which is again model dependent.
In the following we present the two cases separately.

\subsection{Harmonic chain}
\label{subsec-setting-hc}

The harmonic chain is a set of coupled oscillators defined by the Hamiltonian
\eq{
\hat H =
\sum_{n=-\infty}^{+\infty}
%\sum_{i}
\left(
\frac{1}{2m}\, p_n^2+\frac{m\omega^2}{2}\, q_n^2 +\frac{K}{2}(q_{n+1} - q_n)^2
\right)
\label{Hhc}
}
where $m$ is the mass of the oscillators, $K$ is the nearest-neighbour coupling,
while the frequency $\omega$ characterizes the confining potential at each site.
The position and momentum operators satisfy the commutation relations
$\left[ q_n, p_m\right]=i\delta_{n,m}$. The Hamiltonian can be simplified by
the canonical transformation
$p_n \to (m K)^{1/4} \, \hat{p}_n$ and $q_n \to (m K)^{-1/4} \, \hat{q}_n $,
which brings \eqref{Hhc} into
\be
\hat H
= \sqrt{K/m}\,
\sum_{n=-\infty}^{+\infty}
\frac{1}{2}
\left(
\hat{p}_n^2+ \frac{\omega^2}{K/m} \, \hat{q}_n^2 +  (\hat{q}_{n+1} -\hat{q}_n)^2
\right).
\label{Hhc2}
\ee
Note that, working in units of $\hbar =1$, the overall prefactor $\sqrt{K/m}$
has the dimension of energy, thus the transformation corresponds to
working with dimensionless positions, momenta and frequency
measured in units of $\sqrt{K/m}$. For simplicity, in our numerical
calculations we shall set $K=m=1$, which fixes the energy scale and leaves
us with a single parameter $\omega$ to be varied.

The ground state of the harmonic chain can be fully characterized by the
correlation functions of positions and momenta. They can be obtained
by standard procedure, via introducing bosonic creation/annihilation operators
and their Fourier modes, which bring the Hamiltonian into a diagonal form.
The calculation of the correlations is then straightforward and yields
\bea
\label{qqint}
\langle \hat{q}_n \hat{q}_m \rangle
&=&
\int_{-\pi}^{\pi} %\!\!\!
\frac{d q}{4\pi}
\frac{\cos[q (n-m)] }{\sqrt{\omega^2+ 4 \big[ \sin(q/2)\big]^2}}
\\
\rule{0pt}{.9cm}
\label{ppint}
\langle \hat{p}_n \hat{p}_m \rangle
&=&
\int_{-\pi}^{\pi}%\!\!
\frac{d q}{4\pi}
\sqrt{\omega^2+ 4  \big[ \sin(q/2)\big]^2}
\; \cos[q (n-m)] \, .
\eea
The correlation matrices are symmetric and translational invariant,
thus their elements depend only on the distance $r=|m -n|$ between the sites.
Since in our numerical calculations the matrix elements will be needed to a very high
precision, it is useful to have a closed form analytical expression which was reported in
\cite{Botero/Reznik04}
\bea
\label{qq}
& & \hspace{-1.5cm}
\langle \hat{q}_n \hat{q}_{n+r} \rangle
%Q_{i,i+r}
\,=\,
\frac{\kappa^{r+1/2}}{2}\,
%\binom{r-1/2}{r}\,
\frac{\Gamma(r+1/2)}{\Gamma(1/2) \, \Gamma(r+1)} \,
_2F_1 \big(  1/2\, , r+ 1/2\, , r+1 \,,  \kappa^2 \,\big) 
\\
\rule{0pt}{.9cm}
\label{pp}
& & \hspace{-1.5cm}
\langle \hat{p}_n \hat{p}_{n+r} \rangle
%P_{i,i+r}
\,=\,
\frac{\kappa^{r-1/2}}{2}\,
%\binom{r-3/2}{r}\;
\frac{\Gamma(r-1/2)}{\Gamma(-1/2) \, \Gamma(r+1)} \,
_2F_1 \big( - 1/2\, , r- 1/2\, , r+1 \,,  \kappa^2 \,\big) \, .
\eea
Here $_2F_1$ is the ordinary hypergeometric function and the parameter $\kappa$
is defined as
\be
\label{kappa}
\kappa
\equiv
\frac{1}{4} \big( \sqrt{\omega^2+4} - \omega  \,\big)^2 .
\ee
Note that $0<\kappa<1$, and thus the correlations in \eqref{qq} and \eqref{pp}
decay exponentially, with the inverse correlation length given by $\xi^{-1}=-\ln \kappa$.
In particular, $\kappa \to 1$ yields the critical point corresponding to the choice
$\omega \to 0$, where the matrix elements in \eqref{qq} become divergent due
to the zero mode.

In order to construct the entanglement Hamiltonian,
one first introduces the \emph{reduced} correlation matrices $Q$
and $P$ by restricting the indices in \eqref{qqint} and \eqref{ppint}
to the segment $[1,N]$, where the notation $i,j$ will be used.
The single-particle spectrum
in \eqref{ehdiag} is then obtained via the Williamson decomposition
of the block-diagonal matrix $Q \oplus P$, which tells us that
\eq{
(2Q)\, \phi_l = \coth\left(\frac{\varepsilon_l}{2}\right) \psi_l 
\;,
\;\;\qquad\;\;
(2P) \, \psi_l = \coth\left(\frac{\varepsilon_l}{2}\right) \phi_l 
\label{epsw}}
where the vectors $\phi_l$ and $\psi_l$ must satisfy the orthonormality conditions
\be
\label{symp-condition}
\sum_{i=1}^N \phi_l(i) \,\psi_{k}(i) 
= \delta_{l,k} \, , \qquad
\sum_{l=1}^N \phi_l(i) \,\psi_l(j)
= \delta_{i,j} \, .
\ee
The equations (\ref{epsw}) imply the following pair of eigenvalue equations
\eq{
(4 PQ)\, \phi_l = \coth^2\left(\frac{\varepsilon_l}{2}\right) \phi_l 
\;,
\;\;\qquad\;\;
(4 QP) \, \psi_l = \coth^2\left(\frac{\varepsilon_l}{2}\right) \psi_l
\label{epsb}}
meaning that $\phi_l$ and $\psi_l$ are the right and left eigenvectors of
the nonsymmetric matrix $PQ$.

Finally, the entanglement Hamiltonian can be transformed back to the
original position and momentum basis
\be
\label{ent-ham HC}
\mathcal{H}
=
\frac{1}{2}
\sum_{i,j=1}^N \!
\Big(
T_{i,j} \, \hat{p}_i \, \hat{p}_j
+
V_{i,j} \, \hat{q}_i \, \hat{q}_j
\Big)
\ee
where the matrices $T$ and $V$ correspond to the kinetic and potential energy parts.
These matrices can be written respectively as
\cite{Casini/Huerta09, Arias_etal17_1, Arias_etal17_2, Banchi/Braunstein/Pirandola15, DiGiulio/Arias/Tonni19}
\be
\label{ent-ham HC TV}
T_{i,j}= \sum_{l=1}^N\,\psi_l(i)\; \varepsilon_l\; \psi_l(j)\;,
\;\;\qquad\;\;
V_{i,j}= \sum_{l=1}^N\,\phi_l(i)\; \varepsilon_l\; \phi_l(j)
\ee
%\textcolor{blue}{[the matrix relation is $T = \mathcal{V}^{\textrm{t}} \mathcal{E}\mathcal{V}$ 
%and $V = \mathcal{U}^{\textrm{t}} \mathcal{E} \mathcal{U}$]}
in terms of the eigenvectors introduced in (\ref{epsb}).

\subsection{Dimerized hopping model}

Our second model is a fermionic chain with dimerized hopping,
given by the Hamiltonian
\eq{
\hat H = - t\sum_{m=-\infty}^{\infty} \left( \frac{1-\delta}{2} \, c_{2m-1}^\dag c_{2m}^{\phantom{\dag}} +
\frac{1+\delta}{2} \, c_{2m}^\dag c_{2m+1}^{\phantom{\dag}} + \mathrm{h.c.} \right)
\label{hdim}}
where $c^{\dag}_m$ and $c_m$ are now fermionic creation and annihilation
operators, satisfying canonical anticommutation relations
$\{c_m , c_n^{\dag}\}=\delta_{m,n}$. The dimerization is governed by the parameter
$\delta$, where $\delta=0$ corresponds to the homogeneous chain while
$\delta=\pm 1$ is the fully dimerized limit, with every second hopping being zero.
We set the overall hopping amplitude to $t=1$.
The Hamiltonian is two-site shift invariant and can be diagonalized after introducing
Fourier modes on the two sublattices. This leads to a two-band structure of the
dispersion $\omega_q = \pm \sqrt{\cos^2 q+\delta^2 \sin^2 q}$ within a reduced
Brillouin zone $q \in \left[-\pi/2,\pi/2\right]$, with the excitation gap given by $2|\delta|$.

The half-filled ground state can be fully characterized in terms of the fermionic
correlation matrix $\braket{c_{m}^\dag c_{n}^{\phantom{\dag}}}$ which has
a checkerboard structure. In particular, the only nonvanishing matrix elements
beyond the diagonal $\braket{c_{m}^\dag c_{m}}=1/2$ are given by
\eq{
\braket{c_{2m-1}^\dag c_{2n}} = \mathcal{C}_{r}-\delta \, \mathcal{S}_r, \qquad
\braket{c_{2m}^\dag c_{2n+1}} = \mathcal{C}_{r}+\delta \, \mathcal{S}_r
}
where $r=2n+1-2m$ and we defined the integrals
\eq{
\mathcal{C}_r = \int_{-\pi/2}^{\pi/2} \frac{d q}{2\pi}
\frac{\cos qr \cos q}{\sqrt{\cos^2 q+\delta^2 \sin^2 q}} \, , \qquad
\mathcal{S}_r = \int_{-\pi/2}^{\pi/2} \frac{d q}{2\pi}
\frac{\sin qr \sin q}{\sqrt{\cos^2 q+\delta^2 \sin^2 q}} \, .
\label{crsr}}
One can notice that the integrals in \eqref{crsr} have a similar structure as that
in \eqref{qqint} giving the position correlations for the harmonic chain.
Indeed, a closed form expression can also be found for the dimerized chain
and reads for $n\ge m$ \cite{Okamoto88}
\bea
\label{cijdim}
\braket{c_{2m-1}^\dag c_{2n}^{\phantom{\dag}}}=
k^{1/2} \mathcal{J}_{n-m}(k) + k^{-1/2} \mathcal{J}_{n-m+1}(k) 
\\
\braket{c_{2m}^\dag c_{2n+1}^{\phantom{\dag}}}=
k^{-1/2} \mathcal{J}_{n-m}(k) + k^{1/2} \mathcal{J}_{n-m+1}(k)
\eea
where we assumed $\delta>0$ and introduced
\eq{
\mathcal{J}_r(k) = (-1)^r \frac{k^{r+1/2}}{2}\,
%\binom{r-1/2}{r}\,
\frac{\Gamma(r+1/2)}{\Gamma(1/2) \, \Gamma(r+1)} \,
_2F_1 \big(  1/2\, , r+ 1/2\, , r+1 \,,  k^2 \,\big)
\label{tIr}
}
and the parameter $k$ is now given by
\eq{
  k \equiv \frac{1-\delta}{1+\delta} \, .
  \label{defk}
}
Note that the expression in \eqref{tIr} is, up to the alternating factor $(-1)^r$,
exactly the same as the one in \eqref{qq} for the harmonic chain.
The correlations thus depend on the dimerization only via the
parameter $0<k<1$, which is again related to the
correlation length as $\xi^{-1}=-\ln k$.

The single-particle spectrum in \eqref{ehdiag} follows from the
eigenvalues of the reduced correlation matrix $C$ as \cite{Peschel03}
\eq{
(1-2C) \, \phi_l = \tanh\left(\frac{\varepsilon_l}{2}\right) \phi_l
\label{epsf}}
which is the expression analogous to the bosonic case \eqref{epsb}.
Writing the entanglement Hamiltonian in the local fermionic basis
\eq{
  \mathcal{H}=  \sum_{i,j=1}^N \, H_{i,j} \, c^{\dag}_i c_j \,
\label{ent_ham}
}
the matrix $H$ follows as
\eq{
H_{i,j}= \sum_{l=1}^N\,\phi_l(i)\; \varepsilon_l\; \phi_l(j)
\label{Hij}}
where $\phi_l$ is the eigenvector corresponding to $\varepsilon_l$ from \eqref{epsf}.

\section{Half-infinite subsystem\label{sec:halfchain}}
\label{sec-half-infinite}

%%%%%%%%%%%%%%%%%%%%%%%%%%%%%%%%%%%%%%%
 \begin{figure}[t!]
\vspace{-.1cm}
\hspace{-.4cm}
% \begin{center}
\includegraphics[width=1.01\textwidth]{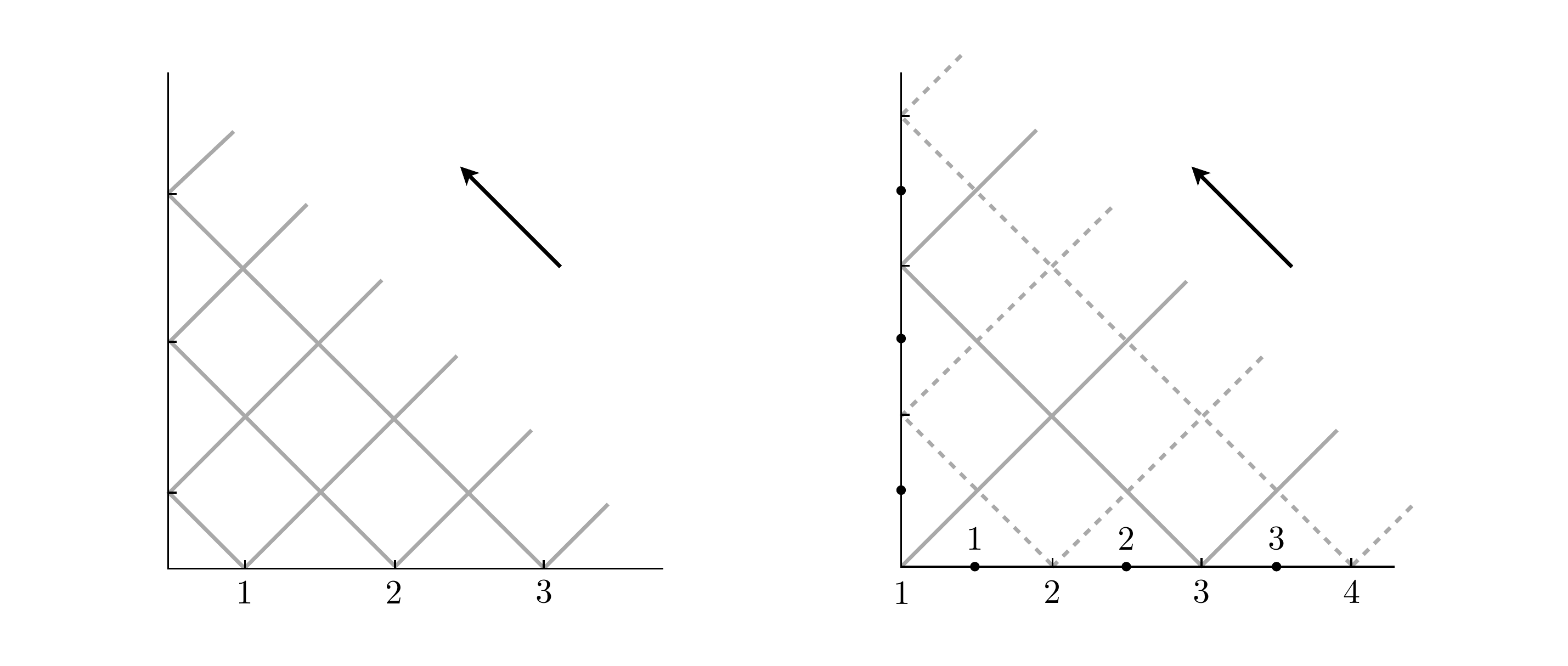}
% \end{center}
\vspace{.1cm}
\caption{
Corner transfer matrix geometries.
Left: Gaussian model related to the oscillator chain.
Right: Interpenetrating Ising models related to the dimerized hopping
chain. Full circles show the location of the dual variables. The
arrows indicate the direction of transfer.
}
\vspace{.0cm}
\label{fig:hc-baxter}
\end{figure}
%%%%%%%%%%%%%%%%%%%%%%%%%%%%%%%%%%%%%%%

In this case, there are explicit expressions for the entanglement Hamiltonians which
result from the relation of the chain problem to an integrable two-dimensional lattice
model and the use of (infinite-size) corner transfer matrices in the latter. This
provides a point of reference for the later treatment of finite subsystems and will
therefore be discussed first.

\subsection{Harmonic chain}
\label{subsec:half-hc}

The harmonic chain can be related to a Gaussian model on a square lattice as described in
\cite{Peschel/Chung99}. The necessary CTM was studied before in \cite{Peschel/Truong91}
and is shown in Fig.\,\ref{fig:hc-baxter} on the left.
This leads to the following expression
for the entanglement Hamiltonian of the half chain with sites $i \ge 1$ if one chooses $m=1$,
$K=\kappa$ and $\omega=1-\kappa$ in \eqref{Hhc}
\be
\label{eh-peschel-chung}
\mathcal{H}_{\textrm{\tiny half}}
\,=\,
2I(\kappa')
\sum_{i=1}^\infty
 \frac{1}{2}
 \Big[\,
 (2i-1)\, p_i^2
+
(2i-1) (1-\kappa)^2\, q_i^2
+
2i\,\kappa \,\big(q_{i+1}-q_i\big)^2
\,\Big]
\ee
where $\kappa' = \sqrt{1-\kappa^2}$ and $I(\kappa)$ is the complete elliptic integral of the first kind which
arises from the elliptic parametrisation of the couplings in the Gaussian model.

To get the result in the parametrisation $K=1$ with $\omega$ being independent,
one needs to carry out the same canonical transformation employed already for the
physical Hamiltonian. In terms of the rescaled variables used in \eqref{Hhc2} one has
\be
\label{ehhchalf}
\mathcal{H}_{\textrm{\tiny half}}
\,=\,
2I(\kappa')\,\sqrt{\kappa}
\sum_{i=1}^\infty
 \frac{1}{2}
 \Big[\,
 (2i-1)\,\hat{p}_i^2
+
(2i-1) \, \omega^2\,\hat{q}_i^2
+
2i \,\big(\hat{q}_{i+1}-\hat{q}_i\big)^2
\,\Big]
\ee
where the rescaled frequency reads
\eq{
\omega^2=(1-\kappa)^2/\kappa \, .
\label{omk}}
Note that since $\omega$ is now the free parameter of the Hamiltonian,
the relation \eqref{omk} must be inverted to get the elliptic parameter
$\kappa(\omega)$. It is easy to see that the solution is given by \eqref{kappa},
such that the elliptic parameter $\kappa$ is identical to the one defining the correlation length.

The operator \eqref{ehhchalf} has thus the same structure as the physical
Hamiltonian, but the coefficients of the terms increase linearly as one moves into the
subsystem. The matrices $T_{i,j}$ and $V_{i,j}$ introduced in \eqref{ent-ham HC}
can be read off the expression, and the only non-zero elements are
\begin{equation}
  T_{i,i}= b(\kappa)\,(2i-1)\,,\quad V_{i,i}= b(\kappa)\,(2i-1)\,(\omega^2 +2)\,,\quad V_{i,i+1}= -b(\kappa)\,2i
  \label{coeff}
\end{equation}
with $b(\kappa)=2I(\kappa')\sqrt{\kappa}$ and $\kappa$ given by \eqref{kappa} in terms of $\omega$.

Finally, the bosonic single-particle eigenvalues $\varepsilon_{l}$ are given by
\cite{Peschel/Chung99}
\begin{equation}
  \varepsilon_{l}= \varepsilon \,(2l-1)\,,\quad \quad \varepsilon = \pi \frac{I(\kappa')}{I(\kappa)}\,,
                     \quad  l=1,2,3, \dots\,.
  \label{eps}
\end{equation}
This result can be checked in the limit $\kappa \rightarrow 0$, where the last term in
\eqref{eh-peschel-chung} vanishes and $\mathcal{H}_{\textrm{\tiny half}}$ becomes the sum of
independent oscillators multiplied by factors $(2i-1)$.

\subsection{Dimerized hopping chain}

The entanglement Hamiltonian for this case has not been given before, but it can be obtained
from known results for the transverse Ising (TI) chain. The reason is that the dimerized chain
is an XX model in spin language which corresponds to two interlacing transverse Ising chains
\cite{Perk/Capel77,Peschel/Schotte84,Turban84,Igloi/Juhasz08}.
Consider the two TI Hamiltonians defined on odd resp. even lattice sites
\begin{eqnarray}
  \hat H_1 &=& - \sum_{m} \left( h_{2m-1} \, \sigma^x_{2m-1} +
                \lambda_{2m-1} \sigma^z_{2m-1} \sigma^z_{2m+1} \right) \\
  \hat H_2 &=& - \sum_{m} \left(  h_{2m} \,\sigma^x_{2m} +
                \lambda_{2m} \,  \sigma^z_{2m} \sigma^z_{2m+2} \right)
\label{TI}
\end{eqnarray}
where $\sigma^x_n, \sigma^z_n$ are Pauli matrices. Then, going over to dual variables
via
\begin{equation}
  \sigma^z_{m} \sigma^z_{m+1}=\tau^z_{m}\, ,\quad \sigma^x_{m}= \tau^x_{m-1}\tau^x_{m}
  \label{dual}
\end{equation}
the total Hamiltonian $\hat H = \hat H_1+\hat H_2$ becomes
\begin{equation}
  \hat H = - \sum_{m} \Big[(h_{2m}\tau^x_{2m-1}\tau^x_{2m}+ \lambda_{2m-1} \tau^z_{2m-1}\tau^z_{2m})
            + (h_{2m+1}\tau^x_{2m}\tau^x_{2m+1}+ \lambda_{2m} \tau^z_{2m}\tau^z_{2m+1})  \Big].
  \label{dualH}
\end{equation}
Therefore one can make the interaction isotropic by choosing
\begin{equation}
 h_{2m}= \lambda_{2m-1},\quad h_{2m+1}= \lambda_{2m}
  \label{parameters}
\end{equation}
which means that the fields in one chain are the couplings in the other one and vice versa.
With a rotation $\tau^z \rightarrow \tau^y$, the Hamiltonian assumes the form
\begin{equation}
  \hat H = - \sum_{m} \Big [\lambda_{2m-1}(\tau^x_{2m-1}\tau^x_{2m}+\tau^y_{2m-1}\tau^y_{2m})
            + \lambda_{2m} (\tau^x_{2m}\tau^x_{2m+1}+ \tau^y_{2m}\tau^y_{2m+1})  \Big ]
  \label{dualH_iso}
\end{equation}
and  describes an inhomogeneous XX chain. The special choice
\begin{equation}
 \lambda_{2m-1}=1-\delta\, , \qquad \lambda_{2m}=1+\delta
  \label{parameters2}
\end{equation}
then leads to the operator \eqref{hdim} if one writes \eqref{dualH_iso} in terms of fermions.
The two TI chains involved are homogeneous but with interchanged parameters.

Now, a \emph{single} TI chain with field $h$ and coupling $\lambda$, is related to an isotropic
two-dimensional Ising model on a square lattice with coupling $K$ if $\lambda/h=\mathrm{sh}^2(2K)$,
and the entanglement Hamiltonian follows from the appropriate CTM as in the bosonic
case \cite{Peschel/Kaulke/Legeza99}. The operator $\mathcal{H}_{\textrm{\tiny half}}$ describes again
a TI chain and differs somewhat for $\lambda/h<1$ (disordered region) and $\lambda/h>1$
(ordered region), see \cite{Davies88}. In the disordered region, it is
\begin{equation}
  \mathcal{H}_{\textrm{\tiny half}} = - \,2I(k')\, \frac{1}{2}\,\sum_{i \ge 1} \Big [ (2i-1) \, \sigma^x_{i}
                                    + k\, 2i \,\sigma^z_{i} \sigma^z_{i+1} \Big ]
  \label{entHTI}
\end{equation}
where $k= \lambda/h$, $k'=\sqrt{1-k^2}$ and $I(k)$ is the same quantity as before.
In the ordered region, $k=h/\lambda$, and $k$ appears in front of the first term in the brackets.

In the present case, one has two interpenetrating Ising lattices, one in the ordered and one
in the disordered region. 
This leads to two interpenetrating CTMs, one
with a tip and one without a tip, as shown in Fig.\,\ref{fig:hc-baxter} on the right,
see also \cite{Truong/Peschel89}.
As a result, the two operators in the exponent satisfy the condition \eqref{parameters} and after
the dual transformation one has
\begin{equation}
  \mathcal{H}_{\textrm{\tiny half}} =  - \,2I(k')\, \frac{1}{2}\, \sum_{i \ge 1}
                                    \Big [\, k\, (2i-1)\, (\tau^x_{2i-1}\tau^x_{2i}+\tau^y_{2i-1}\tau^y_{2i})
                             + 2i\,(\tau^x_{2i}\tau^x_{2i+1}+ \tau^y_{2i}\tau^y_{2i+1})  \Big ]
  \label{entHdim1}
\end{equation}
where now, using \eqref{parameters2}, the parameter $k$ is given by $k=(1-\delta)/(1+\delta)$
as in \eqref{defk}. Writing this in terms of fermions, one arrives at the final result for a half-chain
with sites $i \ge 1$
\begin{equation}
  \mathcal{H}_{\textrm{\tiny half}} =  - \,2I(k') \sum_{i \ge 1}
                                    \Big [\, k\, (2i-1)\, (c_{2i-1}^\dag c_{2i}^{\phantom{\dag}}+ \mathrm{h.c.})
                                   + 2i\, (c_{2i}^\dag c_{2i+1}^{\phantom{\dag}} + \mathrm{h.c.})  \Big ].
  \label{entHdim2}
\end{equation}
This is a hopping model with hopping amplitudes which increase linearly and, in addition,
alternate between 1 and $k$ in exactly the same way as in the physical Hamiltonian \eqref{hdim}
(if one divides $\hat H$ by $(1+\delta)$). Thus the pattern of strong and weak bonds
reappears in the entanglement Hamiltonian, as found in earlier numerical calculations
\cite{Eisler/Chung/Peschel15}. Note that $\mathcal{H}_{\textrm{\tiny half}}$
in \eqref{entHdim2} starts with a weak bond between $i=1$ and $i=2$, i.e. the chain is divided
at a strong bond. If one wants to consider the opposite situation, the factor $k$ has to be moved
to the other term in the bracket.

The fermionic single-particle eigenvalues $\varepsilon_{l}$ of $\mathcal{H}_{\textrm{\tiny half}}$
are given by an expression as for a single TI chain and analogous to the bosonic case
\begin{equation}
  \varepsilon_{l}= \varepsilon \, 2l\,,\quad \quad \varepsilon = \pi \frac{I(k')}{I(k)}\,,
                     \quad  l=0,\pm 1,\pm 2,\pm 3, \dots
  \label{epsfhalf}
\end{equation}
where the factor $2l$ can be checked by taking the limit $k \rightarrow 0$ in
\eqref{entHdim1} or \eqref{entHdim2}. The pairs $(\varepsilon_{l}, \varepsilon_{-l})$ arise
from the two TI operators in the original representation and the state with $l=0$ is the analogue
of the surface state one finds in the Hamiltonian $\hat H$ if the chain is actually cut at the
strong bond.
For a chain divided at a weak bond, one has to move the factor $k$ as mentioned above, and this
changes  $2l$ into  $2l-1$ in the formula.

%%%%%%%%%%%%%%%%%%%%%%%%%%%%%%%%%%%%%%%%%%%%%%%%%%%%
\section{Interval in the harmonic chain\label{sec:osc}}
\label{sec-int-hc}

\begin{figure}[t!]
\vspace{.0cm}
\hspace{-.5cm}
% \begin{center}
\includegraphics[width=\textwidth]{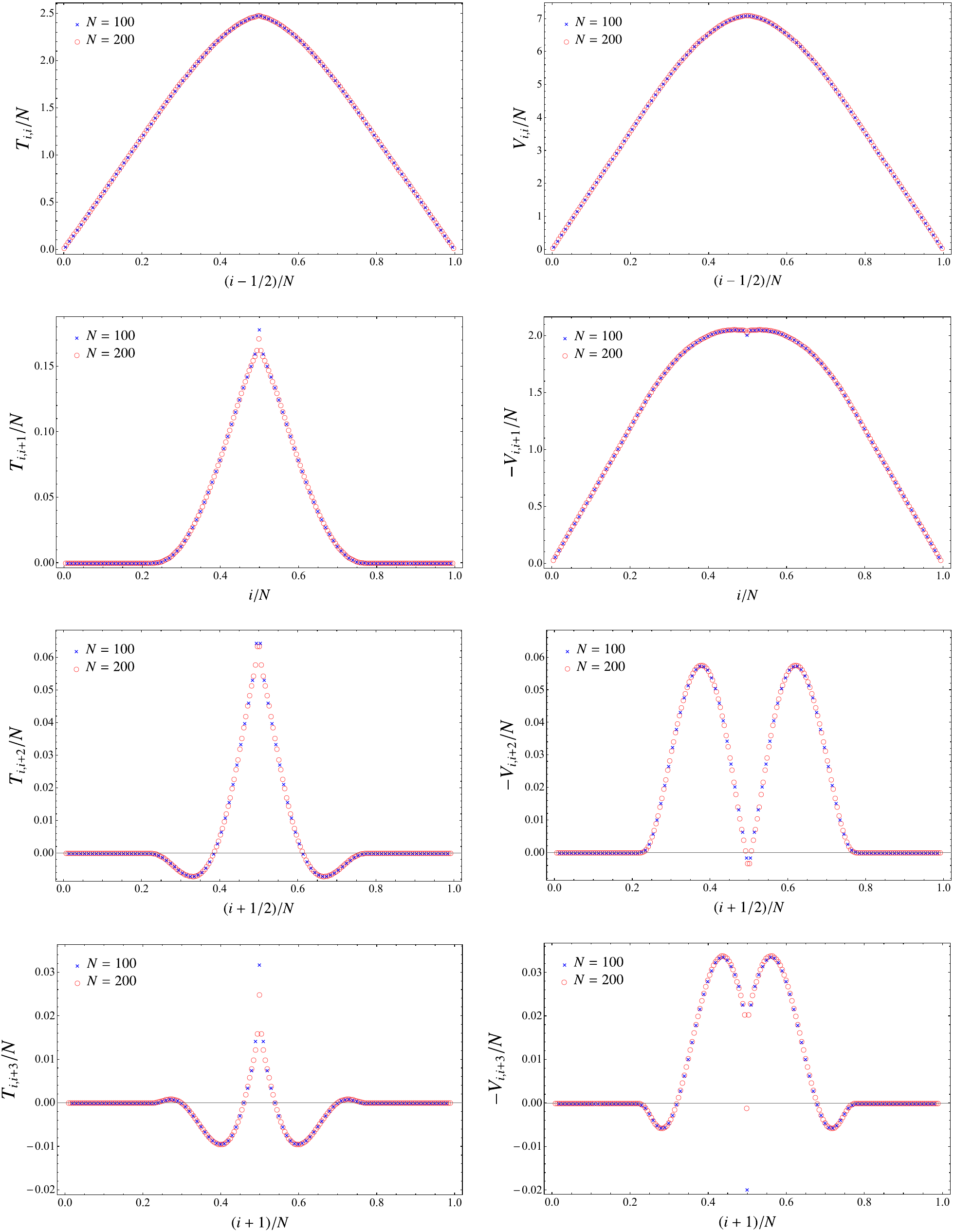}
% \end{center}
\vspace{.2cm}
\caption{
Short-range couplings in the matrices $T$ (left panels) and $V$ (right panels) for $\omega =1$
and two segment sizes. Note the different vertical scales.
}
\vspace{.5cm}
\label{fig:HC-diags}
\end{figure}

\begin{figure}[t!]
\vspace{.0cm}
\hspace{-.4cm}
% \begin{center}
\includegraphics[width=1.0\textwidth]{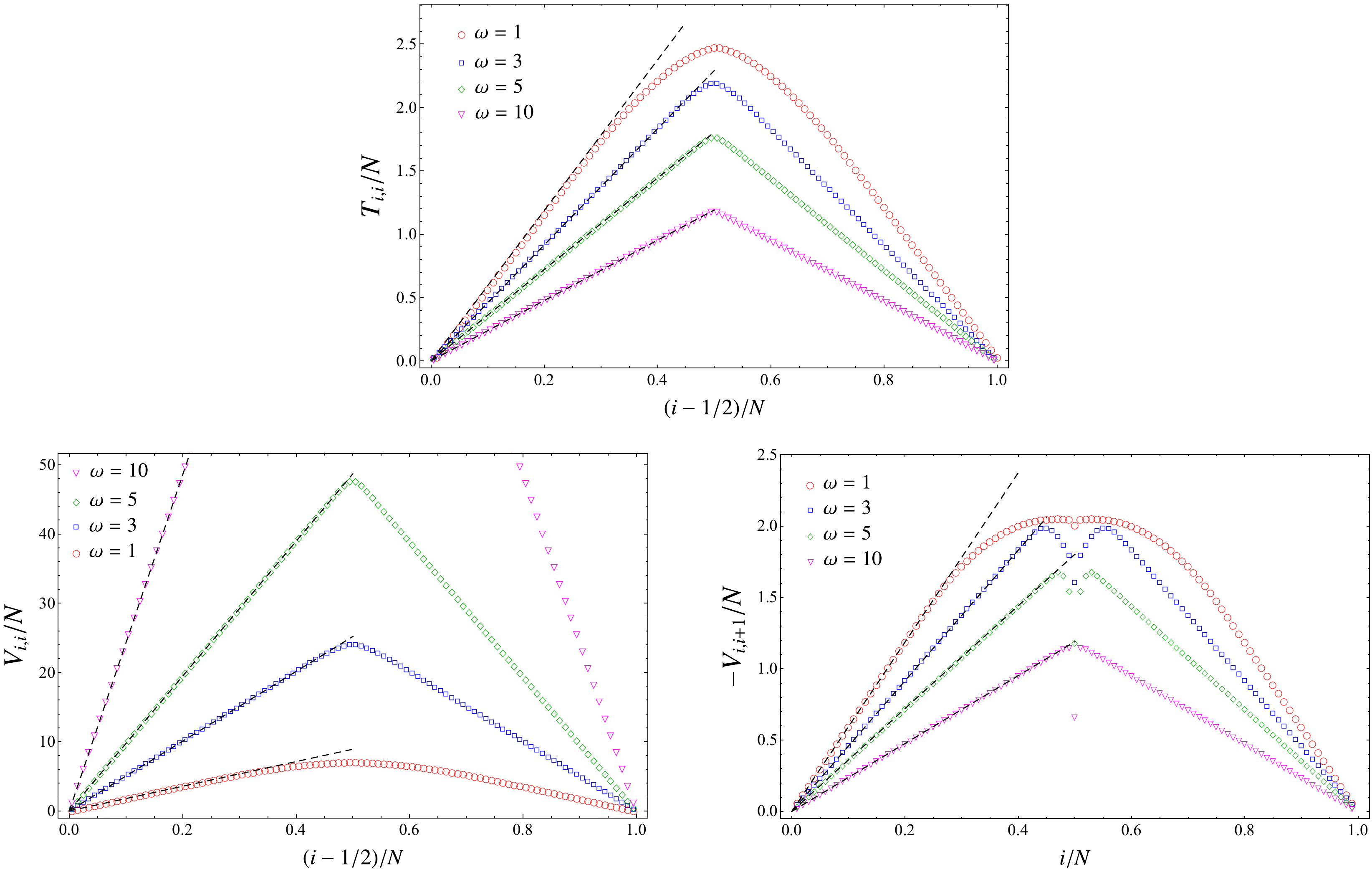}
% \end{center}
%\vspace{-.0cm}
\caption{
Dominant matrix elements of $T$ and $V$ for $N=100$ and various values of $\omega$.
The black dashed lines correspond to the
three-diagonals approximation (\ref{tridiag-EH-singleinterval-T}) and (\ref{tridiag-EH-singleinterval-V}).
}
\vspace{.0cm}
\label{fig:hc-main-diag}
\end{figure}

In this section we consider a finite block made by $N$ consecutive sites in the harmonic chain
and calculate the entanglement Hamiltonian numerically from the correlation matrices 
via (\ref{ent-ham HC TV}).
As mentioned earlier, we set $m=1$ and $K=1$ in (\ref{Hhc}) so that only the oscillator frequency $\omega$ remains, 
from which $\kappa$, related to the correlation length, can be obtained via (\ref{kappa}).
The numerical data shown in Fig.\,\ref{fig:HC-diags}, where $N=200$ and  $\omega=1$, 
have been obtained through a numerical precision given by 800 digits, 
while for Fig.\,\ref{fig:hc-main-diag}, 
where $N=100$ and $\omega=10$, 
we have employed 1000 digits.
In general, we have observed that
higher precision is required as $N$ or $\omega$ increase.

In Fig.\,\ref{fig:HC-diags} the elements in and near the diagonals of the matrices $T$ and $V$ are shown 
for $\omega=1$, which corresponds to $\kappa =0.383$ and a correlation length $\xi =1.04$. 
From previous investigations \cite{DiGiulio/Tonni19} one expects $\mathcal{H}$  to  be extensive, 
therefore the matrix elements are divided by $N$.
Dividing also the site indices by $N$, one finds a perfect collapse of the data for $N=100$ and $N=200$ 
and thus a well-defined limiting behaviour.

In the kinetic energy, only the diagonal elements $T_{i,i}$ are large and show a variation with $i$ which
lies somewhere between a parabola and a triangular form. 
The next elements $T_{i,i+1}$ have a sharp cusp in the middle of the interval and are already 
an order of magnitude smaller. This cusp remains
in the following elements which are still smaller and, in addition, develop more and more structures,
including zeros which do not occur in the case of a critical chain \cite{DiGiulio/Tonni19}.

In the potential energy, the diagonal elements $V_{i,i}$ are again the largest ones, with a shape similar to that of $T_{i,i}$. 
However, here the nearest-neighbour terms $V_{i,i+1}$ are also large, negative and show a kind of plateau in the centre. 
Only the terms describing the interactions with more distant neighbours are much smaller and show structures resembling those in the kinetic terms.
Note that we have plotted $-V_{i,i+r}$ for $r>0$. These are the spring constants if one rewrites the potential
energy properly and therefore typically positive.

A particular feature is  that the structures in the small matrix elements only appear in a certain region
in the centre of the subsystem, while the quantities are zero in the rest of the interval. This region
is the same for all quantities and its width becomes smaller and approaches zero as $\omega$ increases, 
i.e. as the coupling between the oscillators in the chain becomes less important 
(see also Fig.\,\ref{fig:hc-density-plots}).

In Fig.\,\ref{fig:hc-main-diag} we look at the three dominant matrix elements $T_{i,i}$, $V_{i,i}$ and $V_{i,i+1}$ in more detail. They are shown there for relatively large values of $\omega$, 
ranging from $\omega=1$ ($\kappa=0.38$, $\xi=1$) to $\omega =10$ ($\kappa=0.01$, $\xi=0.2$) 
and one sees that all approach a triangular shape as $\omega$ increases.
The dashed lines are the slopes found in (\ref{coeff}) for the half-infinite subsystem and describe
the results very well. This suggests an approximation which consists in keeping only these elements
and setting
\be
\label{tridiag-EH-singleinterval-T}
\frac{T_{i,i}}{N} =
2\,b(\kappa)\, \Delta((i-1/2)/N) 
\ee
and
\be
\label{tridiag-EH-singleinterval-V}
\frac{V_{i,i}}{N} =
%\sqrt{m\lambda}\;
2\big( \omega^2+2\big) \,b(\kappa)\, \Delta((i-1/2)/N)  \,,
\;\qquad\;
 \frac{V_{i,i+1}}{N} =
 -\, 2\,b(\kappa) \, \Delta(i/N) 
\ee
with the ``triangular'' function
\be
\Delta(x) \equiv \frac{1}{2} - \left|\, x  -\frac{1}{2}\, \right|
=
\Bigg\{\begin{array}{ll}
x & 0\leq x \leq 1/2
\\
1-x \hspace{.6cm}& 1/2 \leq x  \leq 1
\\
\end{array}
\label{triangle}
\ee
replacing the simple linear behaviour in (\ref{coeff}). 
In physical terms, this three-diagonals approximation models the entanglement  Hamiltonian $\mathcal{H}$ 
of the interval by glueing the half-infinite ones attached to the endpoints together. 
This should be good for small correlation lengths 
and the analytical expressions allow to predict how the slopes vary with $\omega$. 
Since $\kappa$ decreases as $\omega$ becomes larger, $b(\kappa)$ also decreases 
while $\omega^2 \,b(\kappa)$ increases.

While this approximation describes $\mathcal{H}$  quite well, 
it neglects the structure in the nearest-neighbour coupling $V_{i,i+1}$  in the middle of the subsystem. 
This probably has to be seen together with the features in the small longer-distance couplings.

 \begin{figure}[t!]
\vspace{.0cm}
\hspace{-.2cm}
% \begin{center}
\includegraphics[width=1.0\textwidth]{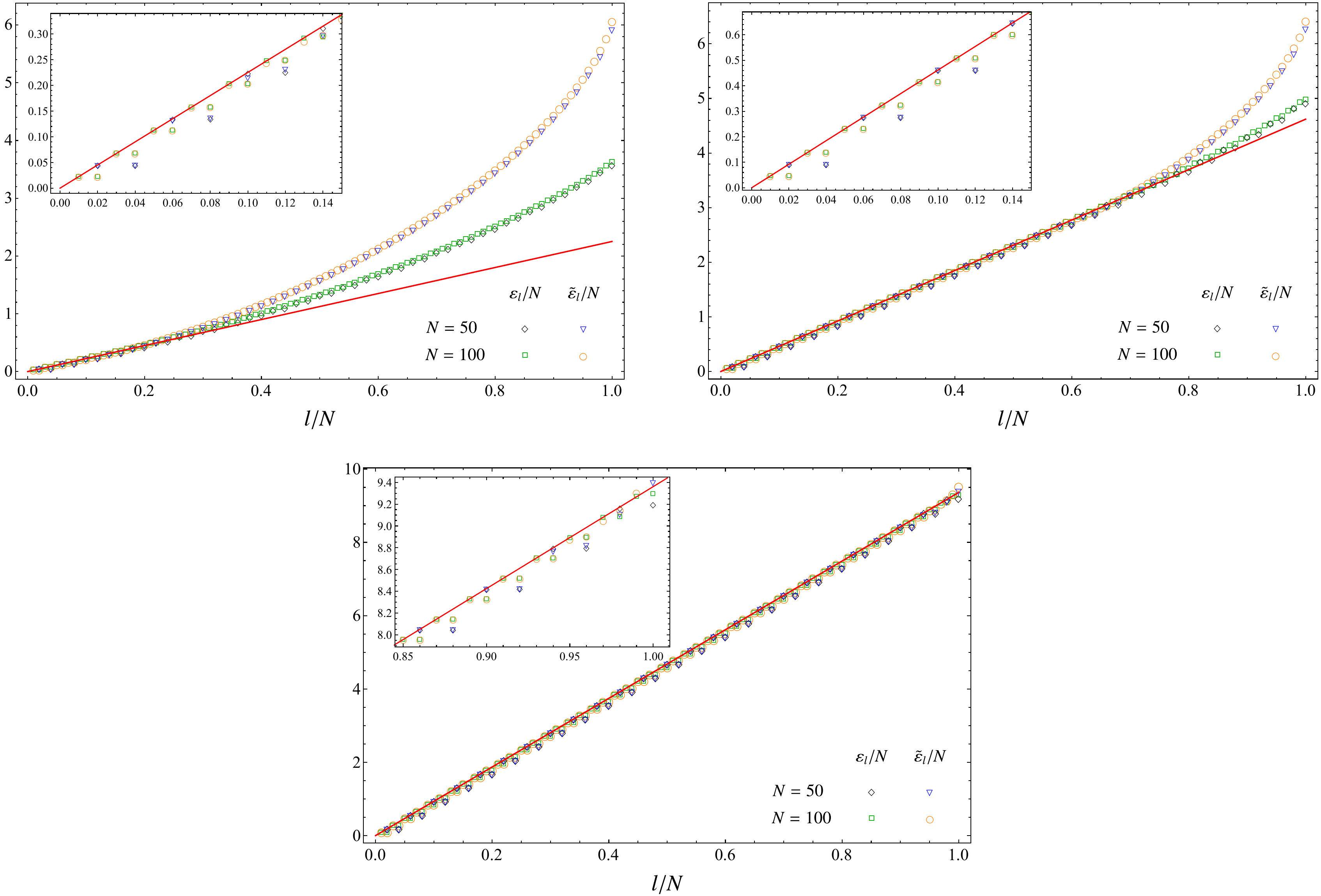}
% \end{center}
%\vspace{-.7cm}
\caption{
Exact and approximate single-particle entanglement eigenvalues
$\varepsilon_l$ and $\tilde{\varepsilon}_l$ for $\omega = 0.1$ (top left),
$\omega = 1$ (top right) and $\omega = 5$ (bottom).
The slope of the red solid line is given by $\varepsilon$ in (\ref{eps}).
}
\vspace{.0cm}
\label{fig:hc-spectrum}
\end{figure}

Finally, we turn to the single-particle spectra $\varepsilon_l$ which follow from the eigenvalues 
of the matrix $PQ$ according to (\ref{epsb}). 
They are shown in Fig.\,\ref{fig:hc-spectrum} for three typical values of $\omega$. 
As $\mathcal{H}$ scales with $N$, so do the $\varepsilon_l$, 
and a plot $\varepsilon_l/N$ vs. $l/N$ gives a universal curve for large $N$. 
Basically, the $\varepsilon_l$ increase at first linearly with $l$, but for large $l$ there is an upward bend. 
This sets in early for small $\omega$ and late for larger $\omega$.
Already for $\omega=5$ the behaviour is just linear. 
The full lines are the results of (\ref{eps}) and are seen to describe the (initial) slope very well. 
A closer look at the smaller eigenvalues for $\omega=0.1$ and $\omega=1$ is provided by the insets and shows that
they are doubly degenerate, as one would expect if one associates them with the two boundaries.
The degenerate levels are described by the half-chain formula \eqref{eps}. 
As the dispersion bends, the degeneracy is also lost.

In Fig.\,\ref{fig:hc-spectrum} the eigenvalues $\tilde{\varepsilon}_l$ result from the entanglement Hamiltonian based on the  three-diagonals approximation 
(\ref{tridiag-EH-singleinterval-T}) and (\ref{tridiag-EH-singleinterval-V}). 
For large $\omega$, a perfect agreement between the two sets is observed
up to the largest few eigenvalues, as shown in the inset for $\omega=5$.
In contrast, for smaller values of $\omega$  the $\tilde{\varepsilon}_l$ lie above the $\varepsilon_l$ at the upper end of the spectrum. 
This is quite reasonable since the triangular form in (\ref{tridiag-EH-singleinterval-T}) and (\ref{tridiag-EH-singleinterval-V})
overestimates the largest matrix elements in the middle of the interval which mainly determine the largest eigenvalues, since the eigenfunctions are concentrated there. 
By contrast, there is always agreement between $\tilde{\varepsilon}_l$ and $\varepsilon_l$ at the lower end.

 \begin{figure}[t!]
\vspace{.2cm}
\hspace{0cm}
\begin{center}
\includegraphics[width=.8\textwidth]{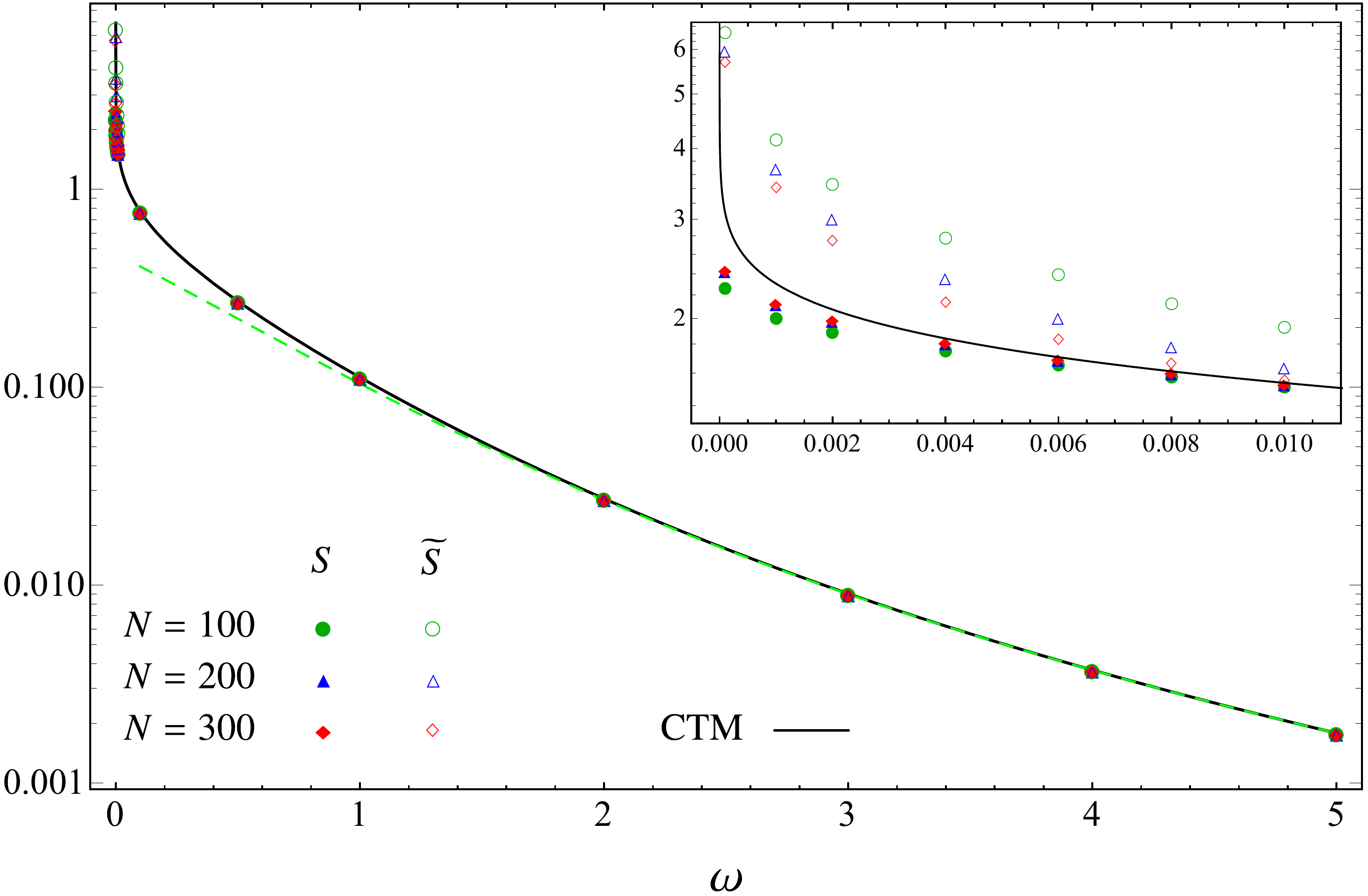}
\end{center}
\vspace{-.0cm}
\caption{
Entanglement entropy as function of $\omega$:
$S$ is obtained through the exact formula (\ref{ee-hc-def}),
$\widetilde{S}$ through the three-diagonals approximation
and the solid black curve corresponds to \eqref{entropyfinal}.
The green dashed line shows the approximation \eqref{entasymp-0} for $\omega \gg 1$,
while the inset shows the behaviour for $\omega \ll 1$.}
\vspace{.0cm}
\label{fig:hc-entropy}
\end{figure}

With the eigenvalues $\varepsilon_l$, the entanglement entropy is given by 
\be
\label{ee-hc-def}
S
=
%\sum_{l=1}^N
%\Big[ (\sigma_l+1/2)\, \log(\sigma_l+1/2) - (\sigma_l-1/2)\, \log(\sigma_l-1/2) \Big]
%=
\sum_{l=1}^N  \left( \frac{\varepsilon_l}{e^{\varepsilon_l} - 1} - \ln \, (1-e^{-\varepsilon_l} ) \right)
\ee
and the result of the numerical calculation is shown in
Fig.\,\ref{fig:hc-entropy} where $S$ is plotted as a function of $\omega$.
For $\omega>1$ one can safely use the spectrum of the half-infinite subsystem 
given in (\ref{eps}) plus the two-fold degeneracy. 
Following the steps sketched in appendix\;\ref{app:ent}, a closed formula for the
entropy can be found as
\begin{equation}
  S= -\frac{1}{12}\left[\ln\left(\frac{16\kappa'^4}{\kappa^2}\right)-(1+\kappa^2)\frac{4I(\kappa)I(\kappa')}{\pi} \right] .
\label{entropyfinal}
\end{equation}
It differs by a factor $2$ from the one reported in \cite{Peschel/Eisler09}
for the half-infinite chain, reflecting the contributions from the two endpoints of the interval.
The result \eqref{entropyfinal} is shown by the solid black line in Fig.\,\ref{fig:hc-entropy}, 
which perfectly agrees with the numerical data.

This agreement actually extends to much smaller $\omega$ as shown in the inset, 
where deviations occur only below $\omega=0.01$ (corresponding to correlation length $\xi=100$). 
The same holds for the entropy $\widetilde{S}$ calculated with the eigenvalues $\tilde{\varepsilon}_l$, 
because $S$ is determined essentially by the low end of the spectrum.

%%%%%%%%%%%%%%%%%%%%%%%%%%%%%%%%%%%%%%%%%%%%%%%%%%%%
\section{Interval in the dimerized chain\label{sec:dim}}

The study of the dimerized hopping chain is somewhat simpler as one has only
the matrix $H$ to consider. The corresponding matrix elements are given by
\eqref{Hij} via the eigenvalue equation \eqref{epsf} of the reduced correlation matrix.
Similarly to the bosonic case, this requires the matrix elements of $C$ to be
calculated with a high precision via the analytic expressions in \eqref{cijdim}-\eqref{tIr}.
Due to the particle-hole symmetry, the nonvanishing entries $H_{i,j}$ are hopping
terms over an odd distance $|j-i|=2p+1$ and it is useful to define their density as
\eq{
h_{i,j} = -\frac{H_{i,j}}{N} \, .
\label{hij}}
%

%%%%%%%%%%%%%%%%%%%%%%%%%%%%%%%%%%%%%%%%%%%%%%%%%%%%%%%%%%%
%
\begin{figure}[t!]
\center
\includegraphics[width=0.49\textwidth]{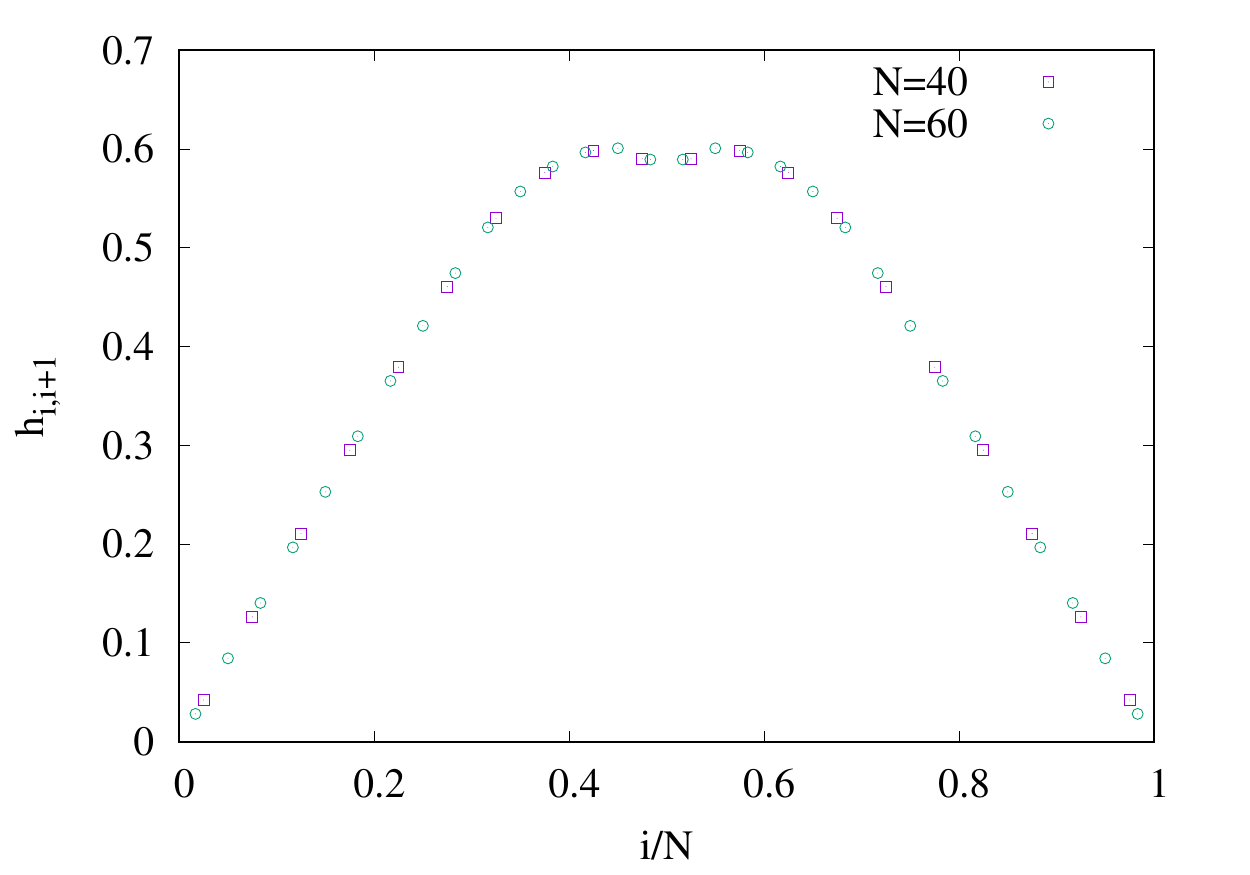}
\includegraphics[width=0.49\textwidth]{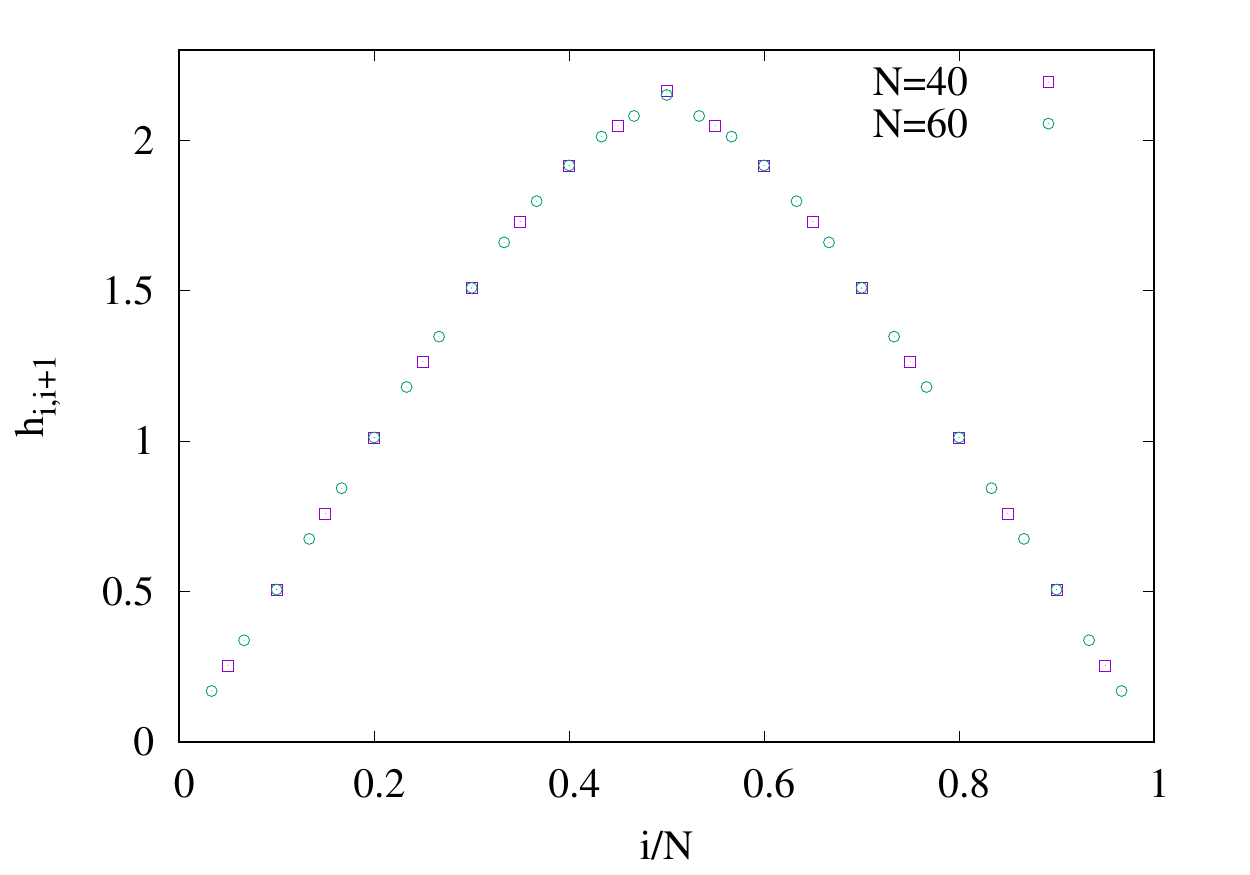}
\includegraphics[width=0.49\textwidth]{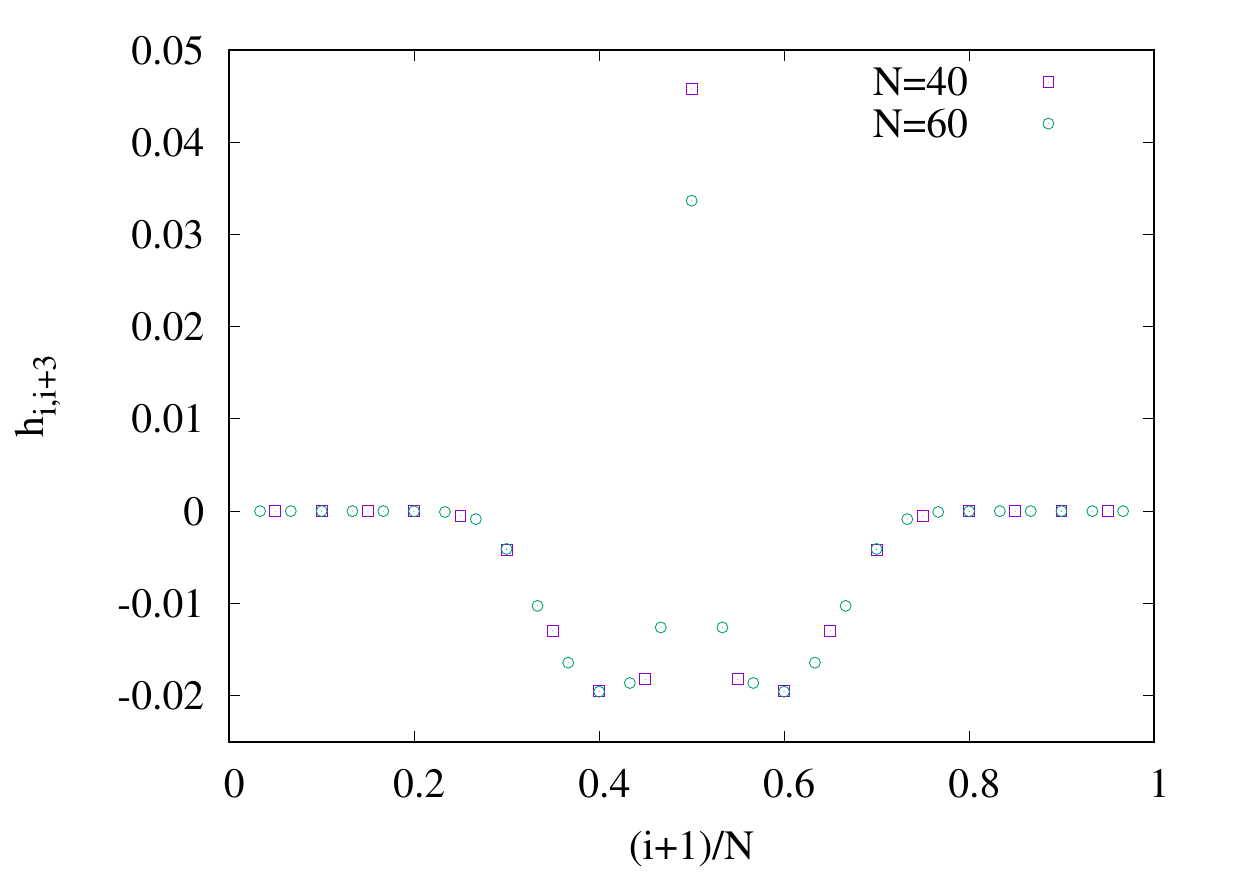}
\includegraphics[width=0.49\textwidth]{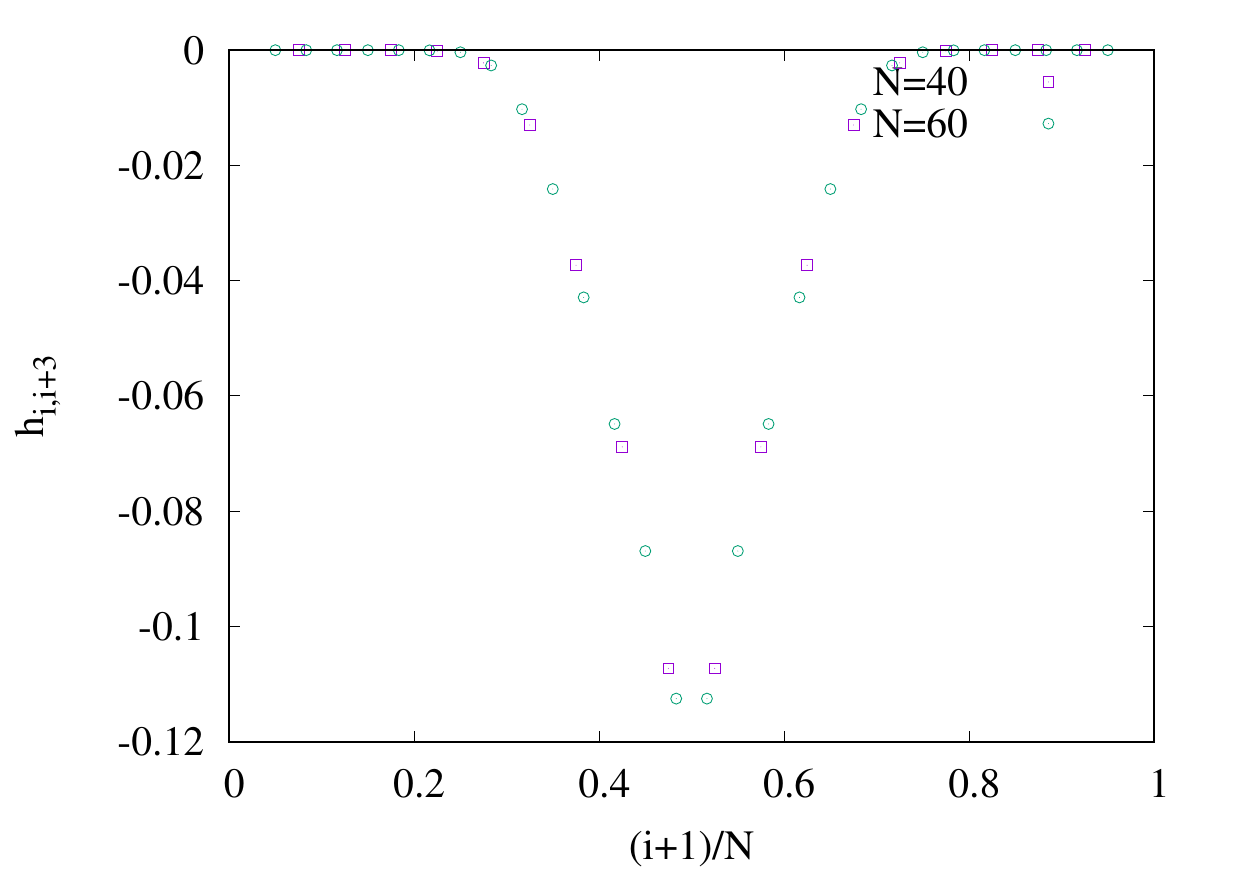}
\includegraphics[width=0.49\textwidth]{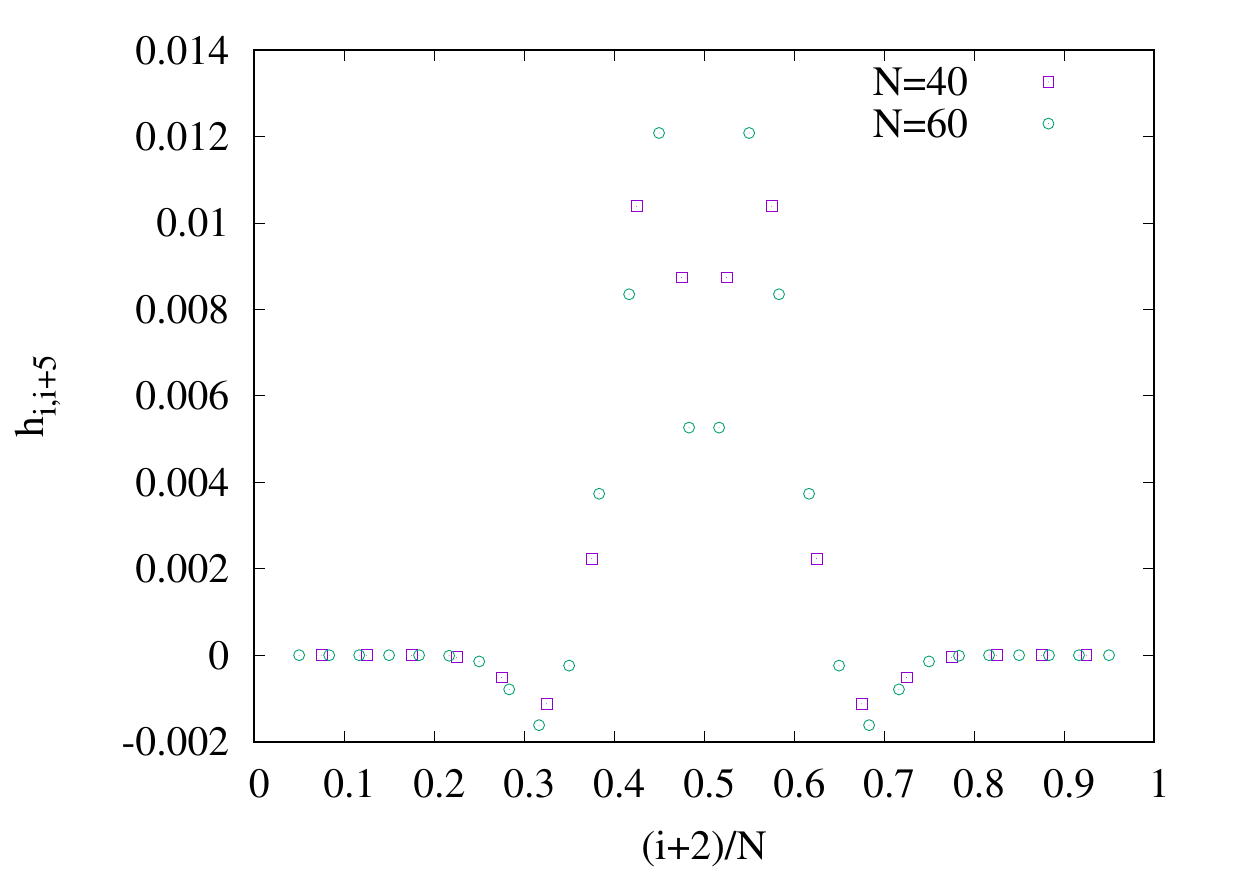}
\includegraphics[width=0.49\textwidth]{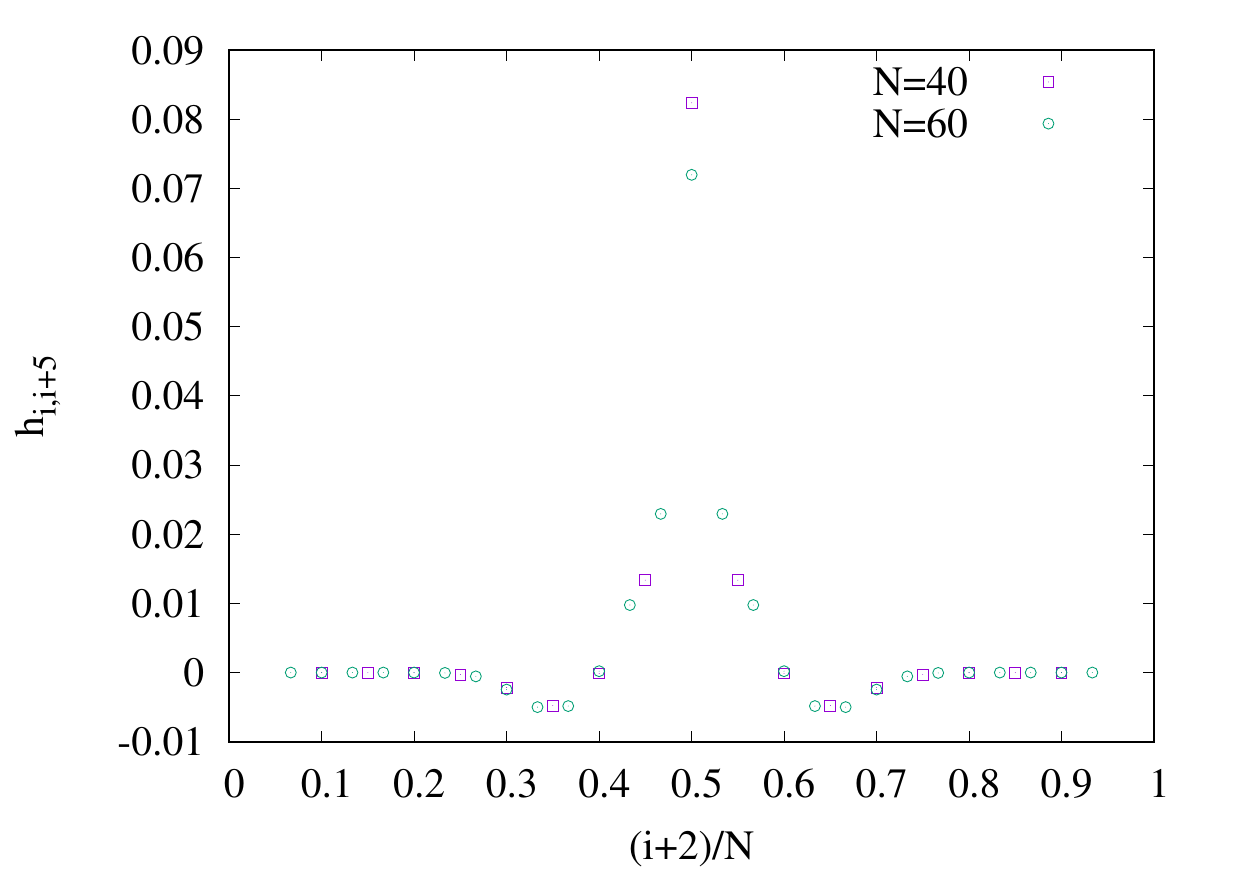}
\caption{First-, third- and fifth-neighbour hopping in $H$ for odd (left)
and even (right) bonds, with dimerization $\delta=0.5$ and for two segment sizes.
Note the different vertical scales.}
\label{fig:hijdim}
\end{figure}
%
%%%%%%%%%%%%%%%%%%%%%%%%%%%%%%%%%%%%%%%%%%%%%%%%%%%%%%%%%%%

To get an overall impression on the structure of the entanglement Hamiltonian,
in Fig.\,\ref{fig:hijdim} we plot the scaled hopping amplitudes in \eqref{hij} along the
diagonals up to the fifth-neighbour terms, for a dimerization parameter $\delta=0.5$.
The hopping amplitudes $h_{i,i+2p+1}$ depend on the scaling variable $(i+p)/N$
as is clear from the data collapse for two different segment sizes.
The hopping matrix is dominated by the nearest-neighbour terms ($p=0$),
similarly to the homogeneous chain $(\delta=0)$. However, the dimerization
induces a strong variation of the hopping across even and odd bonds, shown
by the left  and right columns in Fig.\,\ref{fig:hijdim}. The third- and fifth-neighbour
hopping ($p=1,2$) is an order of magnitude smaller and has a nontrivial structure,
developing sharp peaks in the center, which is reminiscent of the behaviour seen
for the oscillator chain in Fig.\,\ref{fig:HC-diags}. Note also that, in contrast to the homogeneous
case where $h_{i,i+2p+1}>0$ for all $p$, the amplitudes $h_{i,i+3}$ are dominantly
negative for the dimerized case. We checked numerically that this sign change
occurs gradually as one moves towards $\delta \to 0$.

%%%%%%%%%%%%%%%%%%%%%%%%%%%%%%%%%%%%%%%%%%%%%%%%%%%%%%%%%%%
%
\begin{figure}[t]
\center
\includegraphics[width=0.7\textwidth]{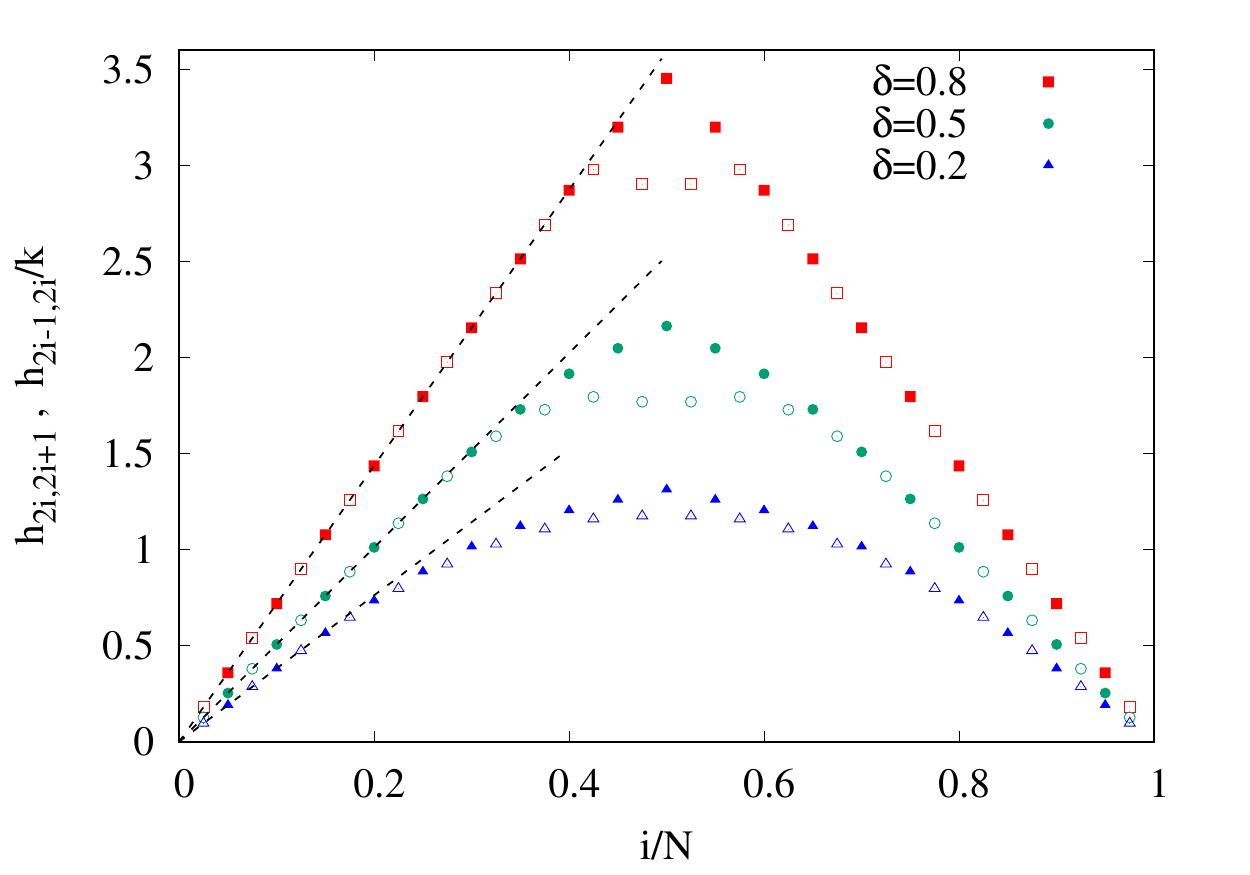}
\caption{Nearest-neighbour hopping in the entanglement Hamiltonian of the dimerized chain
for various $\delta$ and $N=40$. The hopping across odd bonds (empty symbols) is divided by a factor
of $k$. The dashed lines have slopes $2I(k')$ corresponding to the result for the half-infinite
subsystem in \eqref{entHdim2}.}
\label{fig:hnndim}
\end{figure}
%
%%%%%%%%%%%%%%%%%%%%%%%%%%%%%%%%%%%%%%%%%%%%%%%%%%%%%%%%%%%

We shall now focus on the nearest-neighbour hopping and use the exact results
for the half-infinite chain in Sec. \ref{sec:halfchain} to obtain an approximate
understanding for the segment. Our main physical argument is that in a
non-critical system with correlation length $\xi \ll N$, the segment should
effectively behave like a half-infinite system around both of its boundaries.
Hence, the result in Eq. \eqref{entHdim2} predicts a linear increase
of the hopping with a slope $2I(k')$, multiplied by a factor of $1$ or $k$
for the strong (even) and weak (odd) bonds. To check this prediction, we have plotted in
Fig.\,\ref{fig:hnndim} the hopping profiles $h_{2i,2i+1}$ and $h_{2i-1,2i}/k$,
and compared them to the half-infinite result shown by the dashed lines.
The linear approximation works perfectly around the boundary of the segment,
with the agreement improving towards the center for larger $\delta$.
One should remark that all the $\delta$ values in Fig.\,\ref{fig:hnndim}
correspond to very short correlation lengths, in particular one has
$\xi \approx 2.5$ for $\delta=0.2$. Nevertheless, the deviation from the
wedge profile for this value is more pronounced. Clearly, in the limit
$\delta \to 0$ one has to recover the result for the critical case \cite{Eisler/Peschel17},
which is roughly parabolic with a slope $2I(0)=\pi$ at the boundaries.
Note also that the odd hopping profile develops a dip around the center,
in contrast to the even profile which has a marked peak.

%%%%%%%%%%%%%%%%%%%%%%%%%%%%%%%%%%%%%%%%%%%%%%%%%%%%%%%%%%%
%
\begin{figure}[t]
\center
\includegraphics[width=0.6\textwidth]{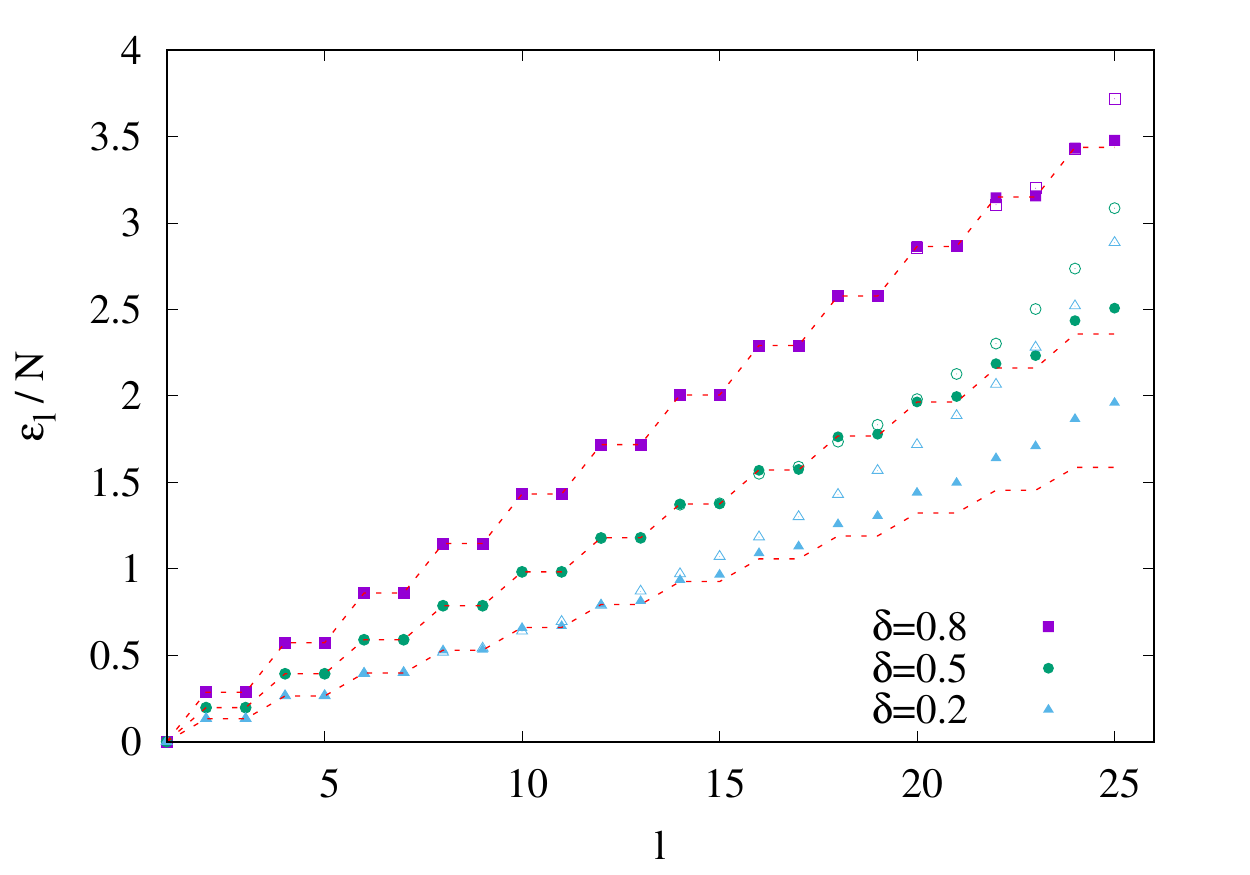}
\caption{Comparison of the scaled single-particle entanglement spectra $\varepsilon_l/N$
(full symbols) to those $\tilde \varepsilon_l/N$ (empty symbols) calculated from $\tilde h$
for $N=50$ and various $\delta$. The dashed red lines show the half-infinite result in
Eq.\,\eqref{epsfhalf} with double degeneracy. Only positive eigenvalues are shown.}
\label{fig:dimspect}
\end{figure}
%
%%%%%%%%%%%%%%%%%%%%%%%%%%%%%%%%%%%%%%%%%%%%%%%%%%%%%%%%%%%

Despite the systematic deviations, one expects that a simple
nearest-neighbour entanglement Hamiltonian $\tilde h_{i,j}$ with
wedge-like hopping amplitudes would 
give a very good approximation $\tilde \rho$ of the actual reduced
density matrix $\rho$. In fact, in the critical case $\delta=0$,
it has recently been shown that such an approximation with a
\emph{parabolic} hopping profile yields a vanishing distance
between $\tilde \rho$ and $\rho$ as $N \to \infty$ \cite{ZCDR20}.
For the dimerized chain we assume, analogously to the oscillator 
chain in \eqref{tridiag-EH-singleinterval-T} and \eqref{tridiag-EH-singleinterval-V}, 
a \emph{triangular}
profile for the nearest-neighbour hopping
\eq{
\tilde h_{2i-1,2i}=2I(k') \, k \, \Delta((2i-1)/N)\,,
\;\;\qquad\;\;
\tilde h_{2i,2i+1}=2I(k')\, \Delta(2i/N)
%w(i) = \frac{1}{2} - \left|\frac{i}{L}-\frac{1}{2}\right|
\label{th}}
where the function $\Delta(x)$ was defined in \eqref{triangle},
and we set $\tilde h_{i,i+2p+1}=0$ for all $i$ and $p>0$.
To check the feasibility of such an approximation, in Fig.\,\ref{fig:dimspect}
we compare the spectra $\tilde \varepsilon_l$ calculated from $\tilde h$
to the actual spectrum $\varepsilon_l$, studied previously in Ref. \cite{Sirker14}.
Note that due to particle-hole symmetry, the eigenvalues come in pairs with
opposite signs, and we show only the positive part of the spectra for better visibility.
Clearly, the low-lying part of the
spectrum is perfectly reproduced, while the larger eigenvalues
$\varepsilon_l$ tend to be overestimated by $\tilde\varepsilon_l$.
The agreement of the high-energy spectrum improves for larger dimerizations,
and for $\delta=0.8$ it already becomes perfect up to the last few eigenvalues.
Note also that the low-lying spectra are doubly degenerate, corresponding to
contributions from the two boundaries, and the levels are given by the CTM result
\eqref{epsfhalf} for the half-infinite chain, shown by the dashed lines in Fig.\,\ref{fig:dimspect}.
The observed features are completely analogous to those shown in
Fig.\,\ref{fig:hc-spectrum} for the oscillator chain.

\begin{figure}[t]
\center
\includegraphics[width=0.6\textwidth]{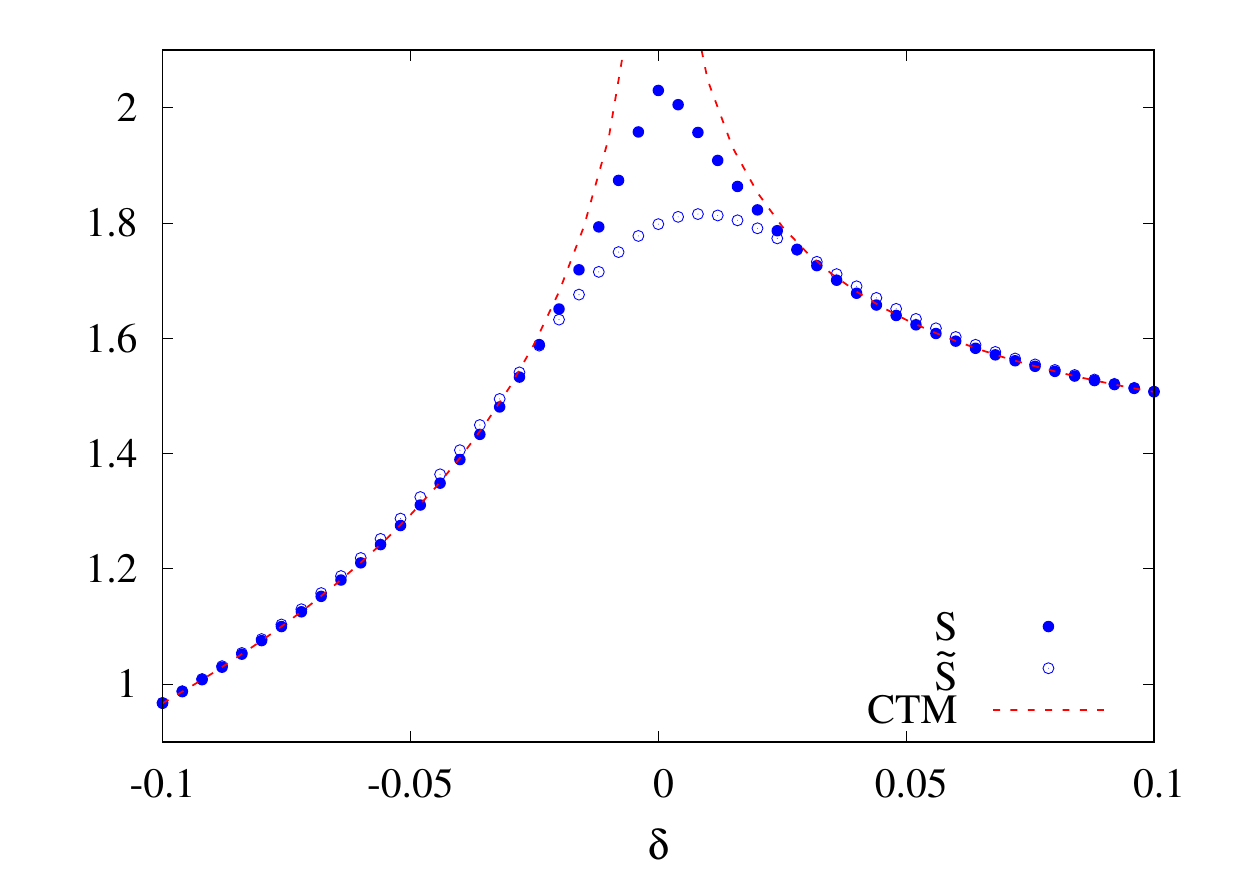}
\caption{Entanglement entropy $S$ and its approximation $\tilde S$ as a function of the
dimerization strength for $N=50$. The red dashed lines show the CTM result in Eq.~\eqref{Sdim}.}
\label{fig:diment}
\end{figure}

It is instructive to have a look also at the entanglement entropy, given by
\be
\label{entdim}
S
=
%\sum_{l=1}^N
%\Big[ (\sigma_l+1/2)\, \log(\sigma_l+1/2) - (\sigma_l-1/2)\, \log(\sigma_l-1/2) \Big]
%=
\sum_{l=1}^N  \left( \frac{\varepsilon_l}{e^{\varepsilon_l} + 1} + \ln \, (1+e^{-\varepsilon_l} ) \right) .
\ee
The quantity $\tilde S$ calculated via $\tilde \varepsilon_l$ is defined analogously.
As only the low-lying $\varepsilon_l$ have a significant contribution,
it is already clear from Fig.\,\ref{fig:dimspect} that $\tilde S$ would give a
perfect approximation of the entropy for the $\delta$ values shown.
Therefore we now focus on smaller dimerizations $|\delta|<0.1$, corresponding to
larger correlation lengths, with the results for $N=50$ shown in Fig.\,\ref{fig:diment}.
Remarkably, the agreement between $S$ and $\tilde S$ remains very good
down to $|\delta| \approx 0.025$ corresponding to $\xi \approx 20$. For even
smaller $|\delta|$ the correlation length exceeds the half-length of the segment,
and the ansatz \eqref{th} built from the contributions of two independent boundaries 
gradually breaks down. The same is true for the doubled CTM result which, using the
formulas for the TI chain \cite{Peschel04b,Peschel/Eisler09}, can be written as
\eq{
S=
\begin{cases}
\displaystyle
\; \frac{1}{3} \left[\ln \Big(\frac{k^2}{16 k'}\Big) + \Big(1 - \frac{k^2}{2}\Big) \frac{4 I(k) I(k')}{\pi}\right] + 2 \ln 2 \;\;\; & \delta>0 
\\
\rule{0pt}{1cm}
\displaystyle
\; \frac{1}{6} \left[\ln \Big(\frac{16}{k^2 k'^2}\Big) + (k^2 - k'^2) \frac{4 I(k) I(k')}{\pi}\right] & \delta<0
\end{cases}
\label{Sdim}}
where for $\delta<0$ one has to use $|\delta|$ in the definition \eqref{defk} such that $k<1$.
In particular, for $\delta \to 0$ ($k\to1$) the CTM result diverges logarithmically.
In contrast, the entropy $\tilde S$ was found to scale as $\tilde S = 1/3 \ln N + \mathrm{const}$,
reproducing the correct prefactor but not the proper constant in $S$. Although the
correct ansatz for the hopping is a parabola for $\delta=0$, the triangular profile has
the same slope at the boundaries and thus reproduces the proper logarithmic scaling of the entropy.

%%%%%%%%%%%%%%%%%%%%%%%%%%%%%%%%%%%%%%%%%%%%%%%%%%%%%%%%%%%
%
%
%%%%%%%%%%%%%%%%%%%%%%%%%%%%%%%%%%%%%%%%%%%%%%%%%%%%%%%%%%%

\section{General features of the non-critical regime\label{sec:genfeat}}

 \begin{figure}[t!]
%\vspace{-.25cm}
%\hspace{-.6cm}
% \begin{center}
\includegraphics[width=\textwidth]{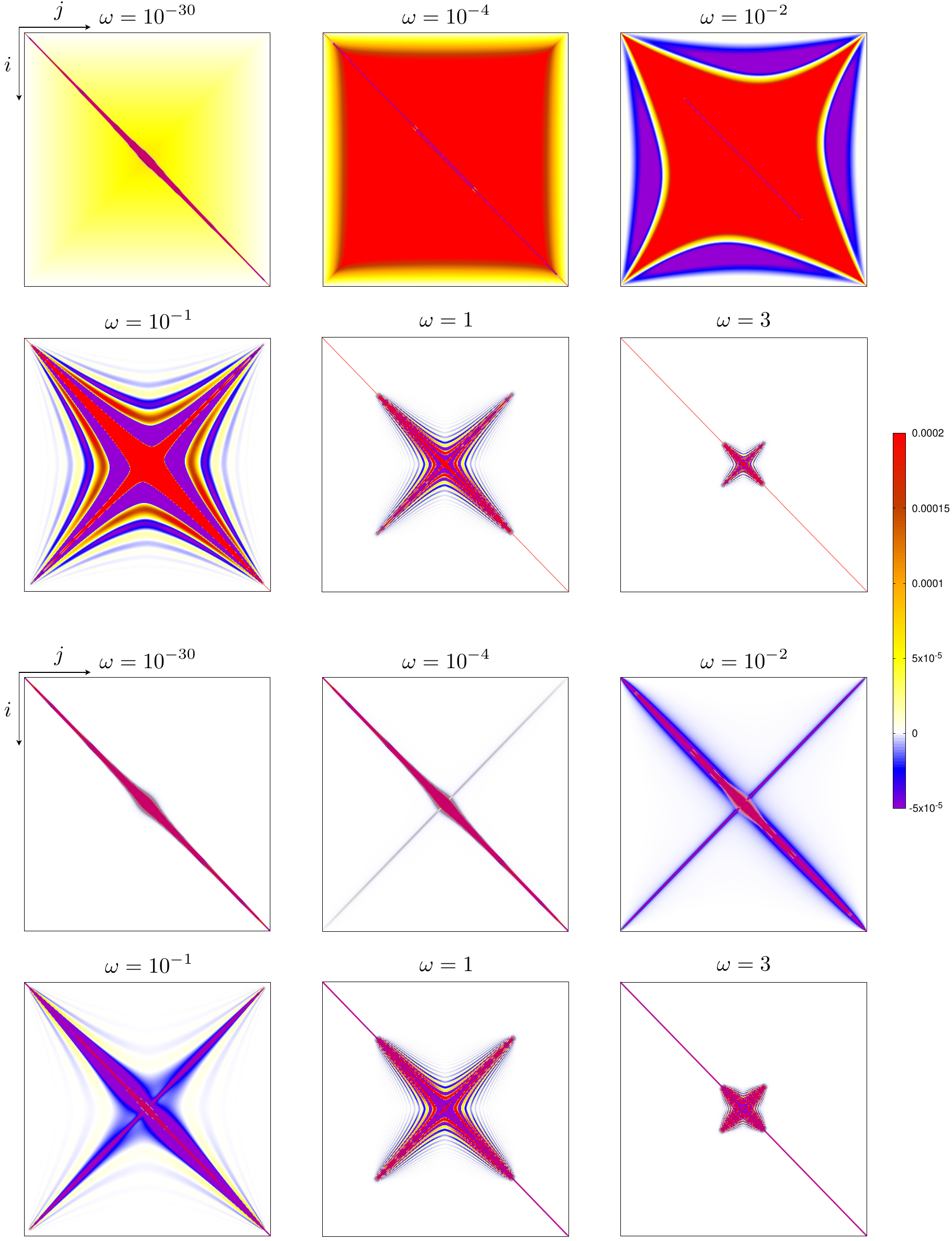}
% \end{center}
\vspace{.5cm}
\caption{
Contour plots of $T_{i,j}/N$ (top)
and $V_{i,j}/N$ (bottom)  for $N=400$ 
and six values of  $\omega$.
%The matrices $T/N$ (top panels) and $V/N$ (bottom panels) 
%for $N=400$ and increasing  $\omega$.
}
\vspace{.0cm}
\label{fig:hc-density-plots}
\end{figure}

In the last two sections, we focussed on strongly non-critical systems with a correlation length
of the order of the lattice constant and thus much smaller than the length of the interval. Here
we want to outline the situation in the whole non-critical region.

For the dominant matrix elements, this was done to some extent already 
in Figs.\,\ref{fig:HC-diags} and \ref{fig:hc-main-diag}
(see Figs.\,\ref{fig:hijdim} and \ref{fig:hnndim} for the fermionic chain), 
where a transition from parabolic to triangular profiles could be observed as $\xi$ became smaller.
The properties of all others are collected in the form of contour plots in 
Fig.\,\ref{fig:hc-density-plots} for the case of the oscillator chain,
where the elements $T_{i,j}/N$ and $V_{i,j}/N$ for $N=400$  and six different values of $\omega$
are shown. 
The size of the elements is given by a colour code where white represents values smaller
than $10^{-5}$. The case $\omega=10^{-30}$ corresponds to a system which is essentially critical and this was studied in detail in \cite{DiGiulio/Tonni19}. 
The finite value of $\omega$ only serves to avoid a zero mode in the chain. 
The cases $\omega=1$ and $\omega=3$ 
correspond to the situation considered in section\;\ref{sec-int-hc}. 
One sees that in both limits the matrices have somewhat larger elements only near the main diagonal. Physically, these are short-range couplings. 
As one moves away from criticality, larger regions of the squares
become filled (in particular for $T_{i,j}$), a cross-shaped structure develops in the middle  and then 
shrinks again. 
Calculations for larger $\omega$ show that it vanishes around $\omega=100$. 
Its finite extent in the direction of the main diagonal was already encountered in 
Fig.\,\ref{fig:HC-diags}, where the matrix elements $V_{i,i+r}$ for small $r$ were seen to vanish beyond a certain distance from the centre.
The elements in the other arm of the cross correspond to longer-range couplings near and across the centre, and a particular case is the sharp ``antidiagonal'' in the matrix of the potential energy, formed by the elements $V_{i,N+1-i}$ which connect points symmetric with respect to the middle of the interval. In particular $i=1$ corresponds to a coupling across the whole subsystem. This structure was already observed in \cite{Arias_etal17_1}.

%%%%%%%%%%%%%%%%%%%%%%%%%%%%%%%%%%%%%%%%%%%%%%%%%%%%%%%%%%%
%
\begin{figure}[t!]
\center
\includegraphics[width=0.7\textwidth]{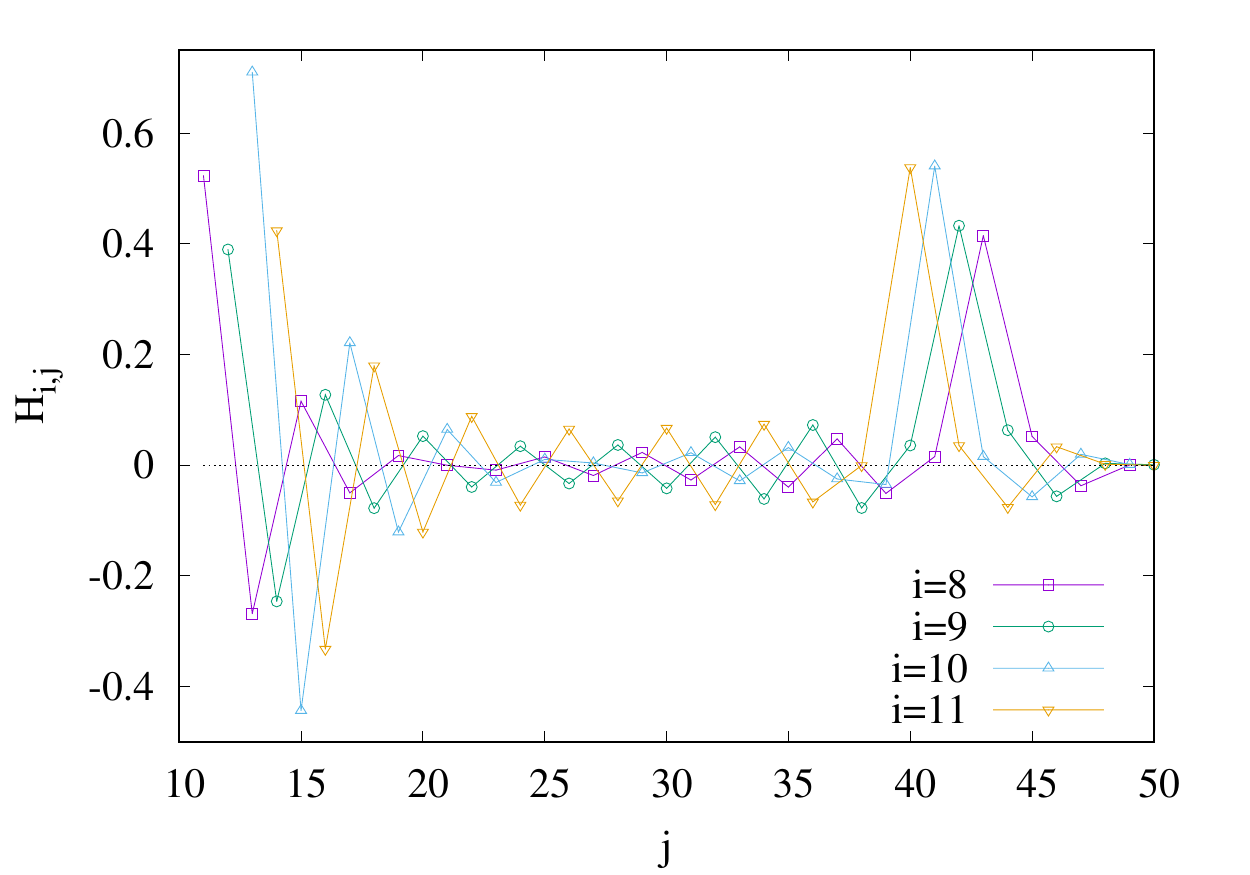}
\caption{Hopping amplitudes $H_{i,j}$ along a fixed row $i$ in the dimerized chain.
The data are shown for $N=50$ as a function of $j>i+1$, omitting the dominant nearest-neighbour term.}
\label{fig:rowcut-fermion}
\end{figure}
%
%%%%%%%%%%%%%%%%%%%%%%%%%%%%%%%%%%%%%%%%%%%%%%%%%%%%%%%%%%%

\begin{figure}[t!]
\center
\includegraphics[width=0.7\textwidth]{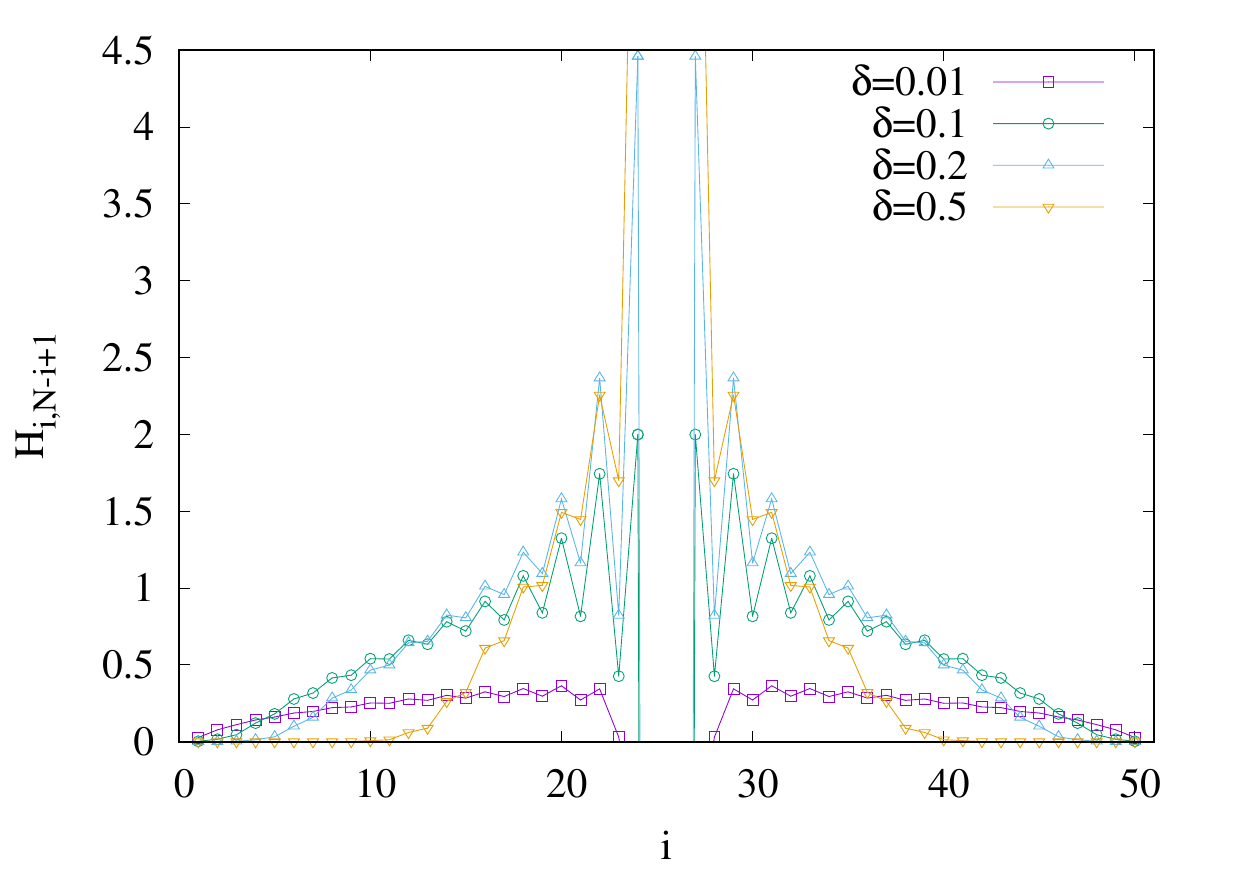}
\caption{Hopping amplitudes $H_{i,N-i+1}$ along the antidiagonal for $N=50$ and various $\delta$,
omitting the dominant nearest-neighbour terms in the middle.}
\label{fig:antidiag-fermion}
\end{figure}

For the dimerized hopping model, an analogous plot of $H_{i,j}/N$ shows similar features and resembles the picture for $\omega=10^{-2}$ in Fig.\,\ref{fig:hc-density-plots}. 
The structure is always cross-like and a sharp antidiagonal exists. 
In Fig.\,\ref{fig:rowcut-fermion}
we present this feature in more detail by showing horizontal cuts through the
matrix, plotting the elements $H_{i,j}$ for fixed $i$ as function of the column index $j$. 
One sees not only a sharp spike right at $j=N+1-i$, 
but already an increase of the values as the antidiagonal is approached
while they are initially decreasing with $j$. 
This behaviour can also be inferred from the contour plots, but is clearer in the direct plot.
As to the values along the antidiagonal, these are shown in Fig. \ref{fig:antidiag-fermion} for
several dimerizations $\delta$. While close to criticality, they are small and decrease only slowly with $i$,
they become larger in the centre for stronger dimerization but also decrease faster, approaching 
zero at some finite point. Remarkably, plotted against $i/N$ and away from the centre, 
the amplitudes $H_{i,N+1-i}$ along the antidiagonal collapse on the same curve  for various $N$  
and are thus nonextensive,
in sharp contrast to the short-range hopping in Fig. \ref{fig:hijdim}.
This is similar to the situation for the central structure in the oscillator chain.
In that case, one finds a similar profile along the antidiagonals,
but the alternations of the dimerized chain are absent.

The phenomenon of the antidiagonals is somewhat intriguing but does not seem to have a simple 
interpretation. In \cite{Arias_etal17_1} it was shown to arise in a perturbative calculation around the critical point, where it comes from the logarithmic oscillations of the critical eigenfunctions, but this is more a
formal argument.

Altogether these results show that the structure of the matrices, as far as the small entries are
concerned, is most complex in the transition region where $\xi \sim N$. This is not unexpected, since there
the effects from both ends of the interval  start to mix, but it will be seen below to cause problems
in a continuum limit.

%%%%%%%%%%%%%%%%%%%%%%%%%%%%%%%%%%%%%%%%%%%%%%%%%%%%%%%%%%%
%
%
%%%%%%%%%%%%%%%%%%%%%%%%%%%%%%%%%%%%%%%%%%%%%%%%%%%%%%%%%%%

\section{Summary and discussion\label{sec:summary}}

We have determined the entanglement Hamiltonian of an interval in a non-critical chain for two
systems which allow for an explicit calculation, one bosonic and one fermionic. In both cases, one had to resort to numerics, but the analytical results for the infinite interval provided a strong
guidance. Quite generally, the matrices describing the quadratic Hamiltonian 
$\mathcal{H}$ in real space contain couplings over arbitrary distances. However, as in the critical cases studied before, only those with short range are large, whereas all others are significantly smaller. In this sense, the situation is simple, and an obvious approximate treatment consists in keeping only the large elements. Using
in addition the analytical results for them then leads to a Hamiltonian with a triangular variation
of the terms along the interval. This was seen to reproduce the low-lying single-particle eigenvalues very accurately over most of the parameter space. As a consequence, also the 
resulting entanglement entropies are correct except in a small region around the critical point.
This is a variant of the ``corner Hamiltonian'' approach 
\cite{Kim/Katsura/Trivedi/Han16, Dalmonte/Vermersch/Zoller18}
in which one replaces the true entanglement Hamiltonian by one with linearly
varying couplings.

All our considerations were for lattice systems, but one can ask about a possible continuum limit 
in the vicinity of the critical point, by introducing a lattice spacing $a$ and taking $a\to 0$.
In fact, for the half-infinite interval this limit can easily be taken
and leads to the Bisognano-Wichmann result (\ref{BW}).
This is outlined in Appendix\;\ref{app:half-line-continuum} for the oscillator chain. 
For the finite interval, one knows that the small longer-range couplings on the 
lattice should be included properly. This leads to sums along horizontal cuts of the corresponding
matrices. For example, the mass parameter $m(x)$ in the continuum description, with $x=ia$,
is given in the oscillator chain by
\be
m(x) = \sum_r V_{i,i+r}     
\ee
whereas the local velocity is
\be
v(x)  = \sum_r r^2 \,V_{i,i+r}
\ee
and a similar expression holds for the local Fermi velocity $v_F(x)$  in the dimerized chain.
It turns out that, in contrast to the situation at criticality, one may need a large number of terms
in order to obtain convergence of the sums, for example 30 terms for $v_F(x)$ if $\delta=0.05$ and $N=100$.
Then $v_F(x)$ shows a triangular profile, but the better converging $m(x)$ looks roughly parabolic with an additional structure in the centre.
However, further increasing the cutoff in the sums, the numerical results for the velocity and mass parameter become unstable,
and even more severe irregularities tend to occur also for the oscillators.
Here, the particular features of the matrices including the antidiagonals enter.
Altogether, we were not able to obtain well-defined general results in the massive regime by fixing $N/\xi$ 
and increasing $N$. This hints toward the possibility that the naive continuum limit,
that perfectly reproduces the CFT results in the massless case \cite{Eisler/Tonni/Peschel19, DiGiulio/Tonni19},
might not be valid away from criticality and that $\mathcal{H}$ remains non-local also in the continuum \cite{Arias_etal17_1}.

A closely related question is, whether a commuting operator with short-range couplings exists in
these non-critical chains. The simple ansatz in Appendix\;\ref{app:comm} 
was not totally successful, but it could be that more general forms like in \cite{Grunbaum/Pacharoni/Zurrian20} do work. 
That would be an important step and would shed additional light on the problem considered here.

\section*{Acknowledgements}

\noindent
ET is grateful to John Cardy and Mihail Mintchev for useful discussions.
VE acknowledges funding from the Austrian Science Fund (FWF) through Project No. P30616-N36.
ET's research has been conducted within the framework of the Trieste Institute for Theoretical Quantum Technologies (TQT).

%\newpage
%%%%%%%%%%%%%%%%%%%%%%%%%%%%%%%%%%%%%%%%%%%%%%%%%%%%%%%%%%%%%%%%%%%%%%%%%%%%%%%%%%%%%%%%%%%%%

\appendix

\section{Half-infinite subsystem: bosonic entanglement entropy}
\label{app:ent}

In this appendix we indicate the steps for obtaining a closed formula for the entanglement entropy 
if the single-particle eigenvalues $\varepsilon_{l}$ of $\mathcal{H}$ 
are given by the CTM result (\ref{eps}) in section \ref{sec-half-infinite}.
The expression (\ref{ee-hc-def}) follows from the general formula
\begin{equation}
  S= \ln Z + U
\label{entropy}
\end{equation}
with the partition function $Z$ and the internal energy $U$. These are given by
\begin{align}
&  1/Z= \prod_{l}(1-e^{-\varepsilon_{l}})= \prod_{{l}=1}^{\infty}(1-q^{2{l}-1})
  \label{Z}
  \\
&  U=\sum_{l} \frac{\varepsilon_{l}}{e^{\varepsilon_{l}}-1}= \varepsilon \sum_{{l}=1}^{\infty}
     (2{l}-1) \frac{q^{2{l}-1}}{1-q^{2{l}-1}}
\label{U}
\end{align}
where $q=\exp(-\pi I(\kappa')/I(\kappa))$.

The product in \eqref{Z} can be obtained from formula (16.37.4) in
\cite{Abramowitz/Stegun64} for the Jacobi theta function $\vartheta_n(u)$ by putting
$m=\kappa^2,m_1= \kappa'^2, u=0$ and using $\vartheta_n(0)=1$. This gives
\begin{equation}
\prod_{{l}=1}^{\infty}(1-q^{2{l}-1})= \left( \frac{16q \kappa'^4}{ \kappa^2} \right)^{1/24}.
\label{prod}
\end{equation}
The sum in \eqref{U} can be obtained from from formula (16.23.10) in
\cite{Abramowitz/Stegun64} for the function ns(u) which reads, correcting a sign,
\begin{equation}
  \mathrm{ns}(u) = \frac{\pi}{2I}\mathrm{csc}(v)+\frac{2\pi}{I} \sum_{{l}=1}^{\infty} \frac{q^{2{l}-1}}
                    {1-q^{2{l}-1}}\sin((2l-1)v) \, , 
                    \quad 
                    v=\frac{\pi u}{2I} \, , \quad I=I(\kappa) \, .
\label{ns}
\end{equation}
Expanding the functions $\mathrm{ns}(u)=1/\mathrm{sn}(u)$ and $\mathrm{csc}(v)=1/\sin(v)$ 
for small $u$ and small $v$ respectively, the leading terms
proportional to $1/u$ cancel and the coefficients of $u$ give the result
\begin{equation}
  \sum_{l=1}^{\infty} (2{l}-1)\frac{q^{2{l}-1}}{1-q^{2{l}-1}}= -\frac{1}{24}\left[1-(1+\kappa^2)
   \left(\frac{2I}{\pi}\right)^2\right] .
\label{sum}
\end{equation}
Taking these results together, one finds for the entropy the formula reported in \cite{Peschel/Eisler09}.

For a comparison with the case of an interval, $S$ should be multiplied by $2$
due to degeneracy of the eigenvalues $\varepsilon_l$, which leads to the result \eqref{entropyfinal}.
%
%\begin{equation}
%  S= -\frac{1}{24}\left[\ln\left(\frac{16\kappa'^4}{\kappa^2}\right)-(1+\kappa^2)\frac{4I(\kappa)I(\kappa')}{\pi} \right]
%\label{entropyfinal}
%\end{equation}
%
For $\kappa\rightarrow 1$, i.e. near criticality, the entropy \eqref{entropyfinal}
diverges and, using $I(\kappa) \simeq \ln(4/\kappa')$, one has for the interval
\begin{equation}
  S \simeq \frac{1}{3}\ln\left(\frac{1}{1-\kappa}\right)
\label{entasymp}
\end{equation}
while for $\kappa \rightarrow 0$ it goes to zero as 
\begin{equation}
S \simeq \frac{1}{4}\,\kappa^2(-\ln\kappa+1/2+\ln 4)
\label{entasymp-0}
\end{equation}
because the coupling of the oscillators vanishes.
The expression \eqref{entasymp-0} in the regime $\omega\gg 1$, that corresponds to
$\kappa\to 0$ from (\ref{kappa}), is shown by the green dashed line in
Fig.\,\ref{fig:hc-entropy}.

\section{Half-infinite subsystem: continuum limit}
\label{app:half-line-continuum}

For a half-infinite interval, the entanglement Hamiltonians in both models have the same structure
as the corresponding chain Hamiltonians. Taking a continuum limit therefore involves the same steps
in both quantities and is rather straightforward. We sketch it here for the oscillator chain.

In place of the discrete variables $\hat q_n$ and $\hat p_n$, fields $\Phi(x)$ and $\Pi(x)$ are introduced via 
\be 
\label{qp-field-replacement}
\hat{q}_n \, \longrightarrow \,  \Phi(x) \, ,
\;\; \qquad \;\;
\hat{p}_n \, \longrightarrow \, a \, \Pi(x)
\ee
where $x=n \, a$ and $a \to 0$ denotes the lattice constant.
Correspondingly, $\Pi(x)$ is a momentum per length. 
Replacing also sums by integrals according to $a \sum_n \to \int d x$
and differences of $\hat q_n$ by first derivatives, 
the Hamiltonian (\ref{Hhc2}) becomes 
\be
\label{HC ham cont}
\hat H
=% \frac{a\, \sqrt{K/m}}{2}
a\, \sqrt{K/m}
 \int_{-\infty}^\infty dx\;
\frac{1}{2} \left[ \,   \Pi(x)^2 + \big(\Phi'(x)\big)^2  + \Omega^2 \,\Phi(x)^2 \,\right]  \, .
%\left(
%\Pi^2+ \Omega^2 \,\Phi^2 +  \big(\Phi'\big)^2
%\right) dx
\ee
Here $\Omega$, which is the mass parameter in the field theory, is given by
\be
\label{Omega-def}
\Omega \equiv \frac{\omega}{a\, \sqrt{K/m}}
\ee
and we recall that $K=m=1$ in our numerical calculations. 
One sees that, if $\Omega$ has to remain finite
for $a \to 0$, also $\omega$ has to vanish in this limit. 
The Hamiltonian (\ref{HC ham cont}) provides the following expression for the energy density in the continuum theory
\be
\label{T00 def}
\mathcal{T}_{00}(x)
\,\equiv\,
\frac{1}{2} \left[ \,   \Pi(x)^2 + \big(\Phi'(x)\big)^2  + \Omega^2 \,\Phi(x)^2 \,\right] .
\ee
\noindent

In the same way, the entanglement Hamiltonian (\ref{ehhchalf}) becomes
\be
\label{Peschel-exact-EH-cont}
\mathcal{H}_{\textrm{\tiny half}}
\,=\,
2 I(\kappa')\,\sqrt{\kappa} 
\int_0^\infty dx\;
 \frac{1}{2}
 \Big[\,
 2x\,\Pi^2
+
2x \, \Omega^2\,\Phi^2
+
2x \,(\Phi')^2
\,\Big] 
\ee
where the factors of $2x$ arise from $(2i-1)a$ and $2ia$, respectively. 
This can be written in terms of the energy density (\ref{T00 def}) as
\be
\label{Peschel-exact-EH-cont2}
\mathcal{H}_{\textrm{\tiny half}}
\,=\,
2b(\kappa)
\int_0^\infty  dx\,
x\,\mathcal{T}_{00}(x) \, ,
\;\;\qquad\;\;
b(\kappa)
\equiv
2 I(\kappa')\,\sqrt{\kappa} \, .
\ee
Here the coefficient $b(\kappa)$ depends on $\omega$ if one uses $\kappa(\omega)$ from (\ref{kappa}).
According to the remark above, $\omega \to 0$ in the continuum limit, which gives 
$\kappa \to 1$ and $\kappa' \to 0$, thus $I(\kappa') \to \pi/2$ 
and one ends up with $2b(\kappa)\to2\pi$, 
which is the value predicted by the 
Bisognano-Wichmann theorem \cite{Bisognano/Wichmann75, Bisognano/Wichmann76}.

\section{Quasi-commuting tridiagonal matrix}
\label{app:comm}

A remarkable feature of the homogeneous hopping chain is the existence of
a tridiagonal matrix that exactly commutes with the matrix $H$ in the entanglement
Hamiltonian \cite{Slepian78,Peschel04}.
For an infinite chain the hopping profile is exactly parabolic, but generalizations to
a finite ring \cite{Grunbaum81} or an open chain \cite{Eisler/Peschel18} also exist. 
Some nontrivial examples of inhomogeneous hopping chains were recently also
uncovered using the theory of bispectrality \cite{CNV19,CNV20}.
Motivated by these examples and the results in section \ref{sec:dim}, a natural
guess of a commuting tridiagonal matrix for the dimerized chain could be given by \\
\eq{
T = %2I(k') \, 
\left(
\begin{array} {ccccc}
0 & t_1   &  &   & \\  
t_1 & 0 & t_2  &  &\\
& t_2 & 0 & t_3  &  \\
& & \ddots & \ddots & \\
& & & t_{N-1} & 0
\end{array} \right)
\label{T}
}
with triangular hopping
\eq{
t_{2i-1}= (1-\delta) \, \Delta \Big(\frac{2i-1}{N} \Big),\qquad
t_{2i}= (1+\delta) \, \Delta \Big(\frac{2i}{N} \Big)
%w(i) = \frac{1}{2} - \left|\frac{i}{L}-\frac{1}{2}\right|
\label{ti}}
where the function $\Delta(x)$ was defined in \eqref{triangle}.

In the following we shall show that, although the matrix $T$
does not exactly commute with $C$ (and hence with $H$), the matrix elements of the
commutator $\left[C,T\right]_{i,j}$ are identically zero for $i,j \le N/2$ and $i,j > N/2$.
Indeed, one has
\eq{
\left[C,T\right]_{i,j} = 
C_{i,j-1} t_{j-1} - C_{i+1,j} t_{i} + C_{i,j+1} t_{j} - C_{i-1,j} t_{i-1}
\label{commct}}
with the boundary conditions $t_0=t_N=0$.
Due to the checkerboard structure of $C$, we only have to consider the cases
$i=2m$ and $j=2n$ or $i=2m-1$ and $j=2n-1$.
Setting $r=2m-2n-1$ and using the definitions \eqref{crsr}, one has for $i,j \le N/2$
\begin{align}
-N \left[C,T\right]_{2m,2n} = \, &\left[ (r+2) \, \mathcal{C}_{r+2} + r \, \mathcal{C}_{r} \right]+
\delta^2 \left [(r+2) \, \mathcal{S}_{r+2} - r \, \mathcal{S}_{r} \right]  \nonumber \\
&+ (2m+2n-1) \, \delta \left[\mathcal{C}_{r+2} - \mathcal{C}_{r} + \mathcal{S}_{r+2} + \mathcal{S}_{r}\right] .
\label{commct2}
\end{align}

Let us first prove that the second line gives zero, i.e. the expression in the
brackets vanishes. This can be proved easily by using only trigonometric identities.
For the piece $\mathcal{C}_{r+2} - \mathcal{C}_{r}$ the trigonometric expression in the numerator
of the integrand becomes
\eq{
(-2 \sin^2 q \cos qr - \sin qr \sin 2q ) \cos q
}
whereas for  $\mathcal{S}_{r+2} + \mathcal{S}_{r}$ one has
\eq{
(2\cos^2 q \sin qr + \sin 2q \cos qr) \sin q \, .
}
One can trivially show that the sum of the two pieces gives zero.

It is more complicated to prove that the first line of \eqref{commct2} also vanishes.
Let us rewrite
\begin{align}
&(r+2) \, \mathcal{C}_{r+2} + r \, \mathcal{C}_{r} =  \int_{-\pi/2}^{\pi/2} \frac{d q}{2\pi}
\frac{\cos q}{\sqrt{\cos^2 q+\delta^2 \sin^2 q}} \frac{d}{d q}(\sin q(r+2) + \sin qr)
\\
&(r+2) \, \mathcal{S}_{r+2} - r \, \mathcal{S}_{r} =  \int_{-\pi/2}^{\pi/2} \frac{d q}{2\pi}
\frac{\sin q}{\sqrt{\cos^2 q+\delta^2 \sin^2 q}} \frac{d}{d q}(-\cos q(r+2) + \cos qr)
\end{align}
and integrate by parts. Using
\begin{align}
 &\frac{d}{d q} \frac{\cos q}{\sqrt{\cos^2 q+\delta^2 \sin^2 q}}=
\frac{ - \delta^2 \sin q}{(\cos^2 q+\delta^2 \sin^2 q)^{3/2}}
 \\
 &\frac{d}{d q} \frac{\sin q}{\sqrt{\cos^2 q+\delta^2 \sin^2 q}}=
\frac{\cos q}{(\cos^2 q+\delta^2 \sin^2 q)^{3/2}}
\end{align}
one can rewrite the term in the first line of \eqref{commct2} as
\eq{
\delta^2
\int_{-\pi/2}^{\pi/2} \frac{d q}{2\pi}
\frac{(\sin q(r+2) + \sin qr)\sin q + (\cos q(r+2) - \cos qr)\cos q}
{(\cos^2 q+\delta^2 \sin^2 q)^{3/2}} \, .
}
Note that the numerator in this integrand is now exactly the same
trigonometric expression that has been shown to vanish above.
Finally, it is also easy to check that the boundary terms from the integration
by parts vanish as well for arbitrary odd $r$.

The calculations for $i=2m-1$ and $j=2n-1$ as well as for the case
$i,j > N/2$ follow similarly.
Unfortunately, however, the absolute value in the expression of the triangular
function $\Delta(x)$ spoils the commutation property if the indices $i$ and $j$ are taken in
different halves of the segment. Nevertheless, for large dimerizations the nonvanishing
matrix elements of the commutator in \eqref{commct} are very small, as the elements
of $C$ decay exponentially with the distance from the diagonal.

\newpage

\section*{References}
%%%%%%%%%%%%%%%%%%%%%%%%%%%%%%%%%%%%%%%%%%%%%%%%%%%%%%%%%%%%%%%%%%%%%%%%%%%%%%%%%

\providecommand{\newblock}{}

\end{document}